\documentclass[journal]{IEEEtranTCOM}
\normalsize
\usepackage[cmex10]{amsmath}
\usepackage{amsfonts}
\usepackage{amsbsy}
\usepackage{amssymb}
\usepackage{array}
\usepackage{mdwtab}
\usepackage{graphicx}
\usepackage{cite}
\usepackage{subfigure}
\usepackage{algorithmic}
\usepackage{cases}
\usepackage{subeqnarray}
\usepackage{multirow}
\usepackage{enumerate}
\usepackage{mathrsfs}
\makeatletter
\def\hlinew#1{
  \noalign{\ifnum0=`}\fi\hrule \@height #1 \futurelet
   \reserved@a\@xhline}
\makeatother
\DeclareMathOperator*{\argmax}{arg\,max}
\def\Var{{\rm Var}\,}
\def\E{{\rm E}\,}
\def\MSE{{\rm MSE}\,}
\def\SNR{{\rm SNR}\,}

\def\AWGN{{\rm AWGN}\,}
\def\MyTheta{\Theta_{q,\boldsymbol\theta}\,}
\def\Aone{({\rm A1})\,}
\def\Atwo{({\rm A2})\,}
\def\Athree{({\rm A3})\,}
\def\Afour{({\rm A4})\,}

\def\Sone{({\rm S1})\,}
\def\Stwo{({\rm S2})\,}
\def\Sthree{({\rm S3})\,}
\def\Sfour{({\rm S4})\,}
\def\Done{({\rm D1})\,}
\def\Dtwo{({\rm D2})\,}
\def\PSK{{\rm PSK}\,}
\def\QAM{{\rm QAM}\,}
\def\FFT{{\rm FFT}\,}

\newcommand{\qed}{\nobreak \ifvmode \relax \else
\ifdim\lastskip<1.5em \hskip-\lastskip
\hskip1.5em plus0em minus0.5em \fi \nobreak
\vrule height0.75em width0.5em depth0.25em\fi}
\title{Accurate Sampling Timing Acquisition for Baseband OFDM Power-line Communication in Non-Gaussian Noise}
\author{Chen~Chen, 
	Yun~Chen,~\IEEEmembership{Member,~IEEE,}
	Na~Ding, 
	Yizhi~Wang,
	Jia-Chin~Lin,~\IEEEmembership{Senior~Member,~IEEE,}
	Xiaoyang~Zeng,~\IEEEmembership{Member,~IEEE,}
	and~Defeng~(David)~Huang,~\IEEEmembership{Senior~Member,~IEEE}%
\thanks{Chen~Chen, Yun~Chen, Na~Ding, Yizhi~Wang, and Xiaoyang~Zeng are with the ASIC and System State Key Laboratory, Fudan University, Shanghai, China (e-mail: \{10212020002, chenyun, xyzeng\}@fudan.edu.cn).}
\thanks{Jia-Chin Lin is with the Department of Communication Engineering, National Central University, Taiwan (e-mail: jiachin@ce.ncu.edu.tw).}
\thanks{Defeng (David) Huang is with the School of Electrical, Electronic and Computer Engineering, The University of Western Australia, Australia (e-mail: david.huang@uwa.edu.au).}}
\begin{document}
\maketitle
\markboth{IEEE Transactions on Communications}%
{Submitted paper}
\begin{abstract} 
In this paper, a novel technique is proposed to address the joint sampling timing acquisition for baseband and broadband power-line communication (BB-PLC) systems using Orthogonal-Frequency-Division-Multiplexing (OFDM), including the sampling phase offset (SPO) and the sampling clock offset (SCO). Under pairwise correlation and joint Gaussian assumption of received signals in frequency domain, an approximated form of the log-likelihood function is derived. Instead of a high complexity two-dimension grid-search on the likelihood function, a five-step method is employed for accurate estimations. Several variants are presented in the same framework with different complexities. Unlike conventional pilot-assisted schemes using the extra phase rotations within one OFDM block, the proposed technique turns to the phase rotations between adjacent OFDM blocks. Analytical expressions of the variances and biases are derived. Extensive simulation results indicate significant performance improvements over conventional schemes. Additionally, effects of several noise models including non-Gaussianity, cyclo-stationarity, and temporal correlation are analyzed and simulated. Robustness of the proposed technique against violation of the joint Gaussian assumption is also verified by simulations.
\end{abstract}

\section{Introduction}
Power-line communication (PLC) is a potential solution for long haul transmission, last mile access, and in-building connection over low to high voltage power networks. By turning virtually each wired device into a target of future value-added services, PLC acts as a technological enabler spanning internet access, residential or business premises, smart grid, and other municipal applications \cite{RecentGalli}. Moreover, it enjoys the advantages of pervasive power cable infrastructures and low implementation cost without rewiring \cite{Meng}.          

To provide a higher data rate than earlier ultra narrowband PLC (UNB-PLC) and narrowband PLC (NB-PLC) solutions, broadband PLC (BB-PLC) empowered by multi-carrier schemes increases the data rate to $200$ Megabits per second (Mbps) and the frequency band to $1.8-30$ MHz \cite{Galli1}. One step further, the new ITU and IEEE standards extend the frequency band to $100$ MHz and increase the data rate to $500$ Mbps \cite{GalliReport}. 

Originally designed for electricity delivery, power line is a harsh and noisy medium, particularly for low power and high frequency data transmission \cite{RecentGalli}. Statistical modeling of PLC channels is a technical challenge for communication theory approach \cite{GalliReport}. In \cite{Zimmermann}, Zimmermann et al. suggested an analytical model for the channel transfer function (CTF) of PLC, characterized by a small number of parameters for frequency below $30$ MHz. To broaden its coverage, Tonello et al. in \cite{Tonello} proposed a random broadband channel generator based on measurement campaign results on nine classes of channels in the $2\sim 100$ MHz range. \cite{Canete,Barmada} discussed the time-variation of PLC channels. However, for broadband transmission, it could be largely neglected, since its period (reciprocal to the mains frequency) is much longer than the block duration. 

Different types of noise are observable in power lines. A previous work \cite{Meng} suggested the Nakagami-m distribution to fit the envelope probability density function (PDF) of time domain background noise. Its accuracy was proved in \cite{tao_statistical_2007} by empirical measurements in a 10-kV medium-voltage power network. Relevant works can be found in \cite{kim_closed-form_2008,BER_kim}. The Class-A impulsive noise model was proposed in \cite{Middleton}. It has been extensively used ever since due to its canonical and closed-form PDF \cite{berry_understanding_1981}, as well as excellent agreement with measurements from both natural and man-made noise environments \cite{Haring}. The power spectral density (PSD) of impulsive noise can reach a value of more than $50$ dB above background noise \cite{Dostert}. In \cite{Katayama}, Katayama et al. revealed the cyclo-stationarity of the noise, synchronous to the mains voltage frequency. Also, noise samples are temporally correlated, as shown in \cite{mo_channel_2009}.

Orthogonal-Frequency-Division-Multiplexing (OFDM) is one option for multi-carrier schemes in BB-PLC. For baseband OFDM systems, signals are transmitted without up-down conversion onto carrier frequency. The complex conjugate property, known as the Hermitian Symmetry Property (HSP) \cite{PolletJour}, must be satisfied in frequency domain after modulation \cite{YongHwa}.

Using Fast Fourier Transform $(\FFT)$, OFDM significantly enhances the system performance in a dispersive propagation environment \cite{BasebandBook}. Nevertheless, it is vulnerable to synchronization imperfections. In this work, two related issues caused by a non-ideal sampling clock at the receiver are considered, including:  
\begin{enumerate}
\item \emph{Sampling phase offset (SPO):} It is caused by the misaligned initial sampling instants between the transmitter and the receiver, resulting in an extra phase rotation proportional to the tone index within one OFDM block (\emph{inter-block phase rotations}) \cite{PolletJour}. It is equivalent to a fractional symbol timing offset (STO) \cite{MIMOBOOK}. Although it could be incorporated into CTF and eliminated by a frequency equalizer (FEQ) \cite{PolletJour}, the residual part should be compensated due to oscillator instabilities \cite{Cortes}.
\item \emph{Sampling clock offset (SCO):} It is caused by a non-synchronized sampling clock between the transmitter and the receiver, triggering not only a inter-block phase rotation, but also a phase growing linearly for successive OFDM blocks (\emph{intra-block phase rotations}) \cite{PolletJour}.
\end{enumerate}

Both SPO and SCO could be tracked recursively using pilots by a delay-locked-loop (DLL) \cite{Byang,Cortes}. The loop operation demands an accurate initial timing acquisition. To the authors' best knowledge, joint sampling timing estimators are seldom discussed in literature. Thus, for comparison, pilot-assisted SPO and SCO estimators are investigated:

\begin{itemize}
\item 
\emph{SPO estimators}: \cite{Coulson} proposed a minimum mean square error (MMSE) estimator for narrowband signals. A low complexity estimator was proposed in \cite{DYK} taking advantage of the phase rotation between two adjacent pilots. \cite{Bo} presented several improved variants to better utilize the phase and power characteristics within one or several OFDM blocks. The scheme in \cite{Ryu} further reduced the pilot distortion using the amplitude information and thus enhanced the performance in a multipath channel. Channel state information (CSI) is explicitly required in \cite{Coulson}, while implicitly in \cite{DYK,Bo,Ryu}. 
\item 
\emph{SCO estimators}: \cite{SpethII} proposed an estimator using the phase rotation between the upper and the lower frequency band without CSI knowledge. In \cite{Morelli}, Morelli et al. proposed a maximum likelihood decoupled estimator (MLDE) and a reduced-complexity estimator (RCE) with superior performance in a multipath channel. CSI could be estimated by the maximum likelihood criterion in \cite{Morelli}. The scheme in \cite{OE} was initially devised for MIMO-OFDM, which was modified to suit a single-antenna system in \cite{Morelli}. \cite{JLS} proposed a weighted joint least square estimator (W-JLSE) outperforming conventional LSE estimators \cite{BasebandBook}. Both \cite{OE} and \cite{JLS} require CSI.      
\end{itemize}

In the present work, a novel joint acquisition technique for SPO and SCO is proposed. An approximated form of the log-likelihood function is derived under pairwise correlation and joint Gaussian assumption of received signals in frequency domain. To avoid the high complexity two-dimension (2-D) grid-search which cannot be decomposed into a one-dimension (1-D) one due to mutual dependency, a five-step method is proposed. First of all, ancillary phases are obtained block-wise, either by a non-data-aided (NDA) estimator or by its data-aided (DA) counterpart. Then, phase unwrapping restores the actual phases from the wrapped ancillary ones. The resultant phases are further embedded into a linear model. With such linearity, LSE is used to separate SPO and SCO. Bias erasure removes the inherent bias of estimation, followed by another LSE which leads to refined estimations. The proposed technique highlights the intra-block rotations with performance improvements over conventional pilot-assisted schemes. Effects of various types of noise and violation of the joint Gaussian assumption are analyzed.   

The remainder of this paper is structured as follows: Section \ref{sec:signalmodel} introduces the signal model. Section \ref{sec:proposed} illustrates the approximated log-likelihood function, several variants of the proposed technique, and analytical expressions for variances and biases. Section \ref{sec:nongaussian} discusses the applicability to time domain non-Gaussianity, cyclo-stationarity, and temporal correlation. Section \ref{sec:sim} presents simulation results. Finally, Section \ref{sec:conclusion} concludes the paper. 

\section{Signal Model}\label{sec:signalmodel}
\footnote{The following notations are used: $(\cdot)^{*}$ as the complex conjugation; $(\cdot)^{T}$ the transposition; $(\cdot)^{H}$ the Hermitian transposition; $\E\{\cdot\}$ the expectation; $\Var\{\cdot\}$ the variance; $\Re\{\cdot\}$ the real part of its argument; $\Im\{\cdot\}$ the imaginary part of its argument; $\mathbb{Z}$ the set of all integers; $\mathbf{I}_{N}$ the $N\times N$ the identity matrix; $\mathbf{J}_{N}$ the $N \times N$ anti-diagonal matrix; $\mathbf{0}_{N_{1}\times N_{2}}$ the $N_{1}\times N_{2}$ all zero matrix; $\mathcal{CN}(\mu(\boldsymbol\mu),\sigma(\boldsymbol\sigma))$ the complex Gaussian distribution with mean $\mu$ (or mean vector $\boldsymbol\mu$) and variance $\sigma$ (or covariance matrix $\boldsymbol\sigma$); $\mathrm{Tr}\{\cdot\}$ the trace of a square matrix; $\max[\cdot]$ the maximum value of the arguments; $\mathrm{diag}\{\mathbf{X}\}$ the diagonal matrix with entries given by vector $\mathbf{X}$; $\boldsymbol\Sigma_{\mathbf{X}}$ the autocorrelation matrix of $\mathbf{X}$ defined as $\E\{\mathbf{X}\mathbf{X}^{H}\}$; $\boldsymbol\Sigma^{p}_{\mathbf{X}}$ the pseudocorrelation matrix of $\mathbf{X}$ defined as $\E\{\mathbf{X}\mathbf{X}^{T}\}$; $[\cdot]_{i,j}$ the $(i,j)$-th entry of a matrix; Other notations would be either self-evident or explained therein.} 
For $Q$ OFDM blocks in frequency domain with $N$ subcarriers in each block, the output signal $s(t)$ at the transmitter on baseband is \cite{Byang}
\begin{equation}
s(t)=\sum_{q=0}^{Q-1}\sum_{k\in\mathcal{K}}X_{q,k}\Psi_{q,k}(t)
\end{equation}
where
\begin{equation}
\Psi_{q,k}(t)=\frac{1}{\sqrt{N}}e^{j2\pi k\left(\frac{t}{NT_{sam}}-\frac{N_{g}+qN_{s}}{N}\right)}u(t-qN_{s}T_{sam})
\end{equation}
is the subcarrier pulse, $\mathcal{K}$ the locations of the data subcarriers, $X_{q,k}$ the equi-probable data modulated on the $k$-th subcarrier in the $q$-th OFDM block and $X_{q,k}\in \mathcal{T}$; $\mathcal{T}$ denotes the finite alphabet of constellation. $N_{s}=N+N_{g}$ is the length of an OFDM block, $N_{g}$ the length of the cyclic prefix (CP), and $T_{sam}$ the sampling interval at the transmitter. The rectangular function $u(t-qN_{s}T_{sam})$ is 
\begin{equation}\label{equ:recfunc}
u(t)=\begin{cases}
1 & 0 \leq t \leq N_{s}T_{sam},\\
0 & \mathrm{otherwise}
\end{cases}
\end{equation}
The multipath channel is 
\begin{equation}\label{equ:channel}
h(t,\tau)=\sum_{l=0}^{L-1}h_{l}(t)\delta(\tau-\tau_{l})
\end{equation}
where $L$ denotes the total number of taps, $\{h_{l}(t)\}_{l=0,1,\cdots,L-1}$ the real channel gains, $\{\tau_{l}\}_{l=0,1,\cdots,L-1}$ the time delay of each path, and $\delta(\cdot)$ the delta function. Transmitting $s(t)$ through the multipath channel \eqref{equ:channel} yields
\begin{equation}
r(t)=s(t)\ast h(t,\tau)+w(t)=\sum_{l=0}^{L-1}h_{l}(t)s(t-\tau_{l})+w(t) 
\end{equation}
where $\ast$ stands for linear convolution, and $w(t)$ is the noise with an unknown distribution and correlation. 

At the receiver, sampling $r(t)$ at time instants $nT_{sam}'$ gives 
\begin{equation}
	r(nT_{sam}')=\sum_{l=0}^{L-1}h_{l}(nT_{sam}')s(nT_{sam}'-\tau_{l})+w(nT_{sam}')
\end{equation}
where $T_{sam}'$ is the sampling interval at the receiver. After removing CP, the $N$ samples in the $q$-th OFDM block is denoted as \begin{math}\mathbf{r}_{q}=[r_{q,0},r_{q,1},\cdots,r_{q,n},\cdots,r_{q,N-1}]\end{math} where \begin{math}
r_{q,n}=r\left((n+qN_{s}+N_{g})T_{sam}'\right)
\end{math}. However, a timing offset $\xi$ exists due to timing imperfections, which consists of the integer part $\mathrm{Int}\{\xi\}$ (caused by the misalignment between the correct starting position of $\FFT$ window and estimated position) and the fractional part $\mathrm{Frac}\{\xi\}$ (caused by the SPO). Without loss of generality, a \emph{left-shift} of $\FFT$ window is assumed with $\mathrm{Int}\{\xi\}\in [-(L-1),0]$. Thus, $r_{q,n}=r\left((n+qN_{s}+N_{g}+\xi)T_{sam}'\right)$.   

\begin{figure*}[tb]
\centering
\includegraphics[scale=0.7]{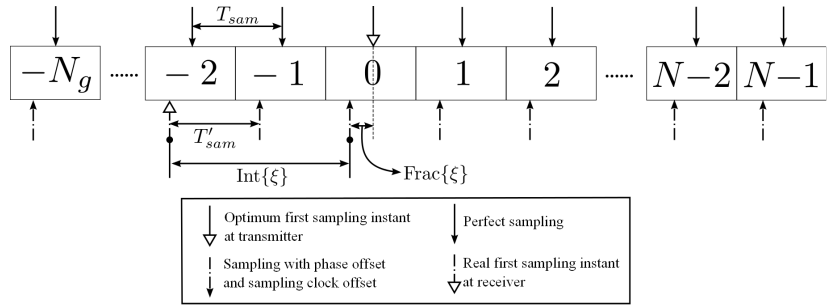}
\caption{Diagram for SPO, SCO, $\mathrm{Int}\{\xi\}$, and $\mathrm{Frac}\{\xi\}$ for the $q$-th OFDM block.}
\label{fig:sampling}
\end{figure*}

The channel is assumed to be quasi-static within one OFDM block's duration. Therefore, $h_{l}(q)$ denotes the real channel gain of the $l$-th path in the $q$-th OFDM block. After demodulation of the $q$-th received OFDM block $\mathbf{r}_{q}$ by $\FFT$, the $k$-th subcarrier is given by \cite{BasebandBook}  
\begin{align}\label{equ:Rfull}
&R_{q,k}=X_{q,k}H_{q,k}e^{j\Theta_{q,k,\xi,\eta}}\Pi(\eta k)\nonumber\\
&+\overbrace{\sum_{\substack{k'=0\\k'\neq k}}^{N-1}X_{q,k'}H_{q,k'}e^{j\Phi(q,k,k',\xi,\eta)}\Pi(k'(1+\eta)-k)}^{\mathrm{ICI}_{\xi,\eta;q,k}}+W_{q,k}
\end{align}
where 
\begin{align}
W_{q,k}&=\sum_{n=0}^{N-1}w\left((n+qN_{s}+N_{g}+\xi)T_{sam}'\right)e^{-j2\pi\frac{kn}{N}}\\
\Pi(\eta k)&=\frac{\sin(\pi\eta k)}{N\sin\left(\frac{\pi\eta k}{N}\right)}\\
\Phi(q,k,k',\xi,\eta)&=\frac{\pi(N-1)(k'(1+\eta)-k)+2\pi(1+\eta)\xi k'}{N}\nonumber\\
&+\frac{2\pi qN_{s}\eta k'+2\pi N_{g}\eta k'}{N}\\
\Theta_{q,k,\xi,\eta}&=\Phi(q,k,k,\xi,\eta)
\end{align}
Here, $H_{q,k}=\sum_{l=0}^{L-1}h_{l}(q)e^{-j2\pi\frac{k\tau_{l}}{NT_{sam}}}$ is the CTF of channel \cite{Speth}, and $\eta=(T_{sam}'-T_{sam})/T_{sam}$ the SCO. The diagram illustrating SPO and SCO is drawn in Fig.~\ref{fig:sampling}. The extra phase rotation $\Theta_{q,k,\xi,\eta}$ can be approximated by 
\begin{equation}
\Theta_{q,k,\xi,\eta}\approx\frac{2\pi \xi k}{N}+\pi \eta k +\frac{2\pi(qN_{s}+N_{g})\eta k}{N}
\end{equation}
for large $N$ and small $\eta$. $\Pi(\eta k)$ is the amplitude attenuation which approaches $1$ for small $\eta$ \cite{Speth}. $\mathrm{ICI}_{\xi,\eta;q,k}$ is the inter-carrier interference (ICI) caused by $\xi$ and $\eta$.   

For convenience, $\mathrm{ICI}_{\xi,\eta;q,k}$ is neglected; its effect is demonstrated in Section \ref{sec:sim}. Thus, \eqref{equ:Rfull} can be written into a compact matrix form below:
\begin{equation}
\mathbf{R}_{q}=\boldsymbol\Phi_{q}\mathbf{H}_{q}\mathbf{X}_{q}+\mathbf{W}_{q}\label{equ:Rsimp}
\end{equation}
where
\begin{math}
	\mathbf{R}_{q}\triangleq[R_{q,0}\ R_{q,1}\ \cdots\ R_{q,N-1}]^{T},\ \boldsymbol\Phi_{q}\triangleq\mathrm{diag}\{e^{j\Theta_{q,k,\xi,\eta}},k\in[0,N-1]\},\ \mathbf{H}_{q}\triangleq\mathrm{diag}\{{{\mathbf{\dot{H}}}_{q}}\},\ {\mathbf{\dot{H}}_{q}}\triangleq[H_{q,0}\ H_{q,1}\ \cdots\ H_{q,N-1}]^{T},\ \mathbf{X}_{q}\triangleq[X_{q,0}\ X_{q,1}\ \cdots\ X_{q,N-1}]^{T},\ \mathbf{W}_{q}\triangleq[W_{q,0}\ W_{q,1}\ \cdots\ W_{q,N-1}]^{T}
\end{math}. Here, we make two assumptions:

\begin{description}
\item[$\Aone$] $\mathbf{H}_{q}\mathbf{X}_{q}\sim\mathcal{CN}(\mathbf{0}_{N\times 1},\sigma_{X}^{2}\sigma_{H}^{2}\mathbf{I}_{N})$.  
\item[$\Atwo$] $\mathbf{W}_{q}\sim\mathcal{CN}(\mathbf{0}_{N\times 1},\sigma_{W}^{2}\mathbf{I}_{N})$ and independent from $\mathbf{H}_{q}\mathbf{X}_{q}$.
\end{description}
Therefore, $\mathbf{R}_{q}$ is a \emph{joint Gaussian vector} distributed as $\mathcal{CN}(\mathbf{0}_{N\times 1},(\sigma_{X}^{2}\sigma_{H}^{2}+\sigma_{W}^{2})\mathbf{I}_{N})$. On the other hand, the \emph{pseudocorrelation matrix} (PCM) of $\mathbf{H}_{q}\mathbf{X}_{q}$ and $\mathbf{W}_{q}$ can be written into  
\begin{align}
&[\boldsymbol\Sigma^{p}_{\mathbf{HX}}]_{i,j}=\begin{cases}
\sigma_{X}^{2}\sigma_{H}^{2},& i+j=N\\ 
0,&\mathrm{otherwise}
\end{cases}\label{equ:pseudoHX}\\
&[\boldsymbol\Sigma^{p}_{\mathbf{W}}]_{i,j}=\begin{cases}
\sigma_{W}^{2},& i+j=N\\ 
0,&\mathrm{otherwise}
\end{cases}\label{equ:pseudoW}
\end{align}
using the proprieties and HSPs of $\mathbf{H}_{q}\mathbf{X}_{q}$ and $\mathbf{W}_{q}$ respectively. For instance, the propriety of $W_{q,k}$ indicates that $\E\{W_{q,k}W_{q,k'}\}=0$ if $k\neq N-k'$, while the HSP implies that $W_{q,k}=W^{*}_{q,N-k}$ and so $\E\{W_{q,k}W_{q,k'}\}=\sigma_{W}^{2}$ \emph{if and only if} $k=N-k'$. 

\emph{Remarks:}\\
In practice, neither $X_{q,k}$ nor $H_{q,k}$ is Gaussian distributed: $X_{q,k}$ is taken from a finite alphabet with equal probability and thus is uniformly distributed, while the statistical model in \cite{Tonello} is adopted for ${H}_{q,k}$. The closed-form distribution of $\mathbf{H}_{q}\mathbf{X}_{q}$ is hard to pursue in practice. Even when both $X_{q,k}$ and $H_{q,k}$ are Gaussian distributed, their product $H_{q,k}X_{q,k}$ follows the Gaussian product distribution which is non-Gaussian. $\Aone$ and $\Atwo$ significantly simplify our analysis by making $R_{q,k}\sim\mathcal{CN}(0,\sigma_{X}^{2}\sigma_{H}^{2}+\sigma_{W}^{2})$. Robustness of the proposed schemes against non-Gaussianity of $\mathbf{R}_{q}$ is demonstrated via simulations in Section \ref{sec:sim}.


\section{Proposed estimators}\label{sec:proposed}
\subsection{Approximated Log-likelihood Function}
An OFDM system with $N_{n}$ null subcarriers at both ends of the spectrum as the guard bands is considered here. The \emph{universal set} containing all the $N$ subcarriers is denoted as $\mathcal{I}_{U}$, which can be decomposed into three subsets with index given by 
\begin{itemize}
\item $\mathcal{I}_{0}=\{k|k=0,k=N/2,k\in \mathbb{Z}\}=\mathcal{I}_{0}^{+}\cup \mathcal{I}_{0}^{-}$ 
\item $\mathcal{I}_{1}=\{k|k=(N-N_{n})/2+v,0\leq v \leq (N_{n}/2)-1,v\in \mathbb{Z}\}\cup\{k|k=(N/2)+v,1\leq v \leq (N_{n}+1)/2+1,v\in \mathbb{Z}\}=\mathcal{I}_{1}^{+}\cup \mathcal{I}_{1}^{-}$
\item $\mathcal{I}_{2}=\{k|1\leq k \leq (N-N_{n})/2-1,k\in \mathbb{Z}\}\cup\{k|(N-N_{n})/2+2 \leq k \leq N-1,k\in \mathbb{Z}\}=\mathcal{I}_{2}^{+}\cup \mathcal{I}_{2}^{-}$ 
\end{itemize}
\begin{figure*}[tb]
\centering
\includegraphics[scale=0.55]{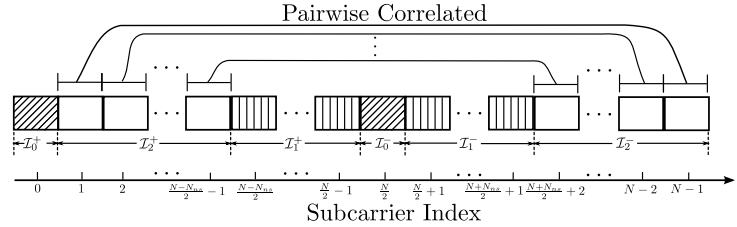}
\caption{Decomposition into $\mathcal{I}_{0}=\mathcal{I}_{0}^{+}\cup\mathcal{I}_{0}^{-},\ \mathcal{I}_{1}=\mathcal{I}_{1}^{+}\cup\mathcal{I}_{1}^{-}, \text{and}\ \mathcal{I}_{2}=\mathcal{I}_{2}^{+}\cup\mathcal{I}_{2}^{-}$.}
\label{fig:hsp}
\end{figure*}
For BB-PLC baseband system, $\mathcal{I}_{0}$ delivers random \emph{real-valued} data, which are often nullified since direct current (DC) and Nyquist frequency subcarriers are generally discarded. $\mathcal{I}_{1}$ contains the null subcarriers, while subcarriers in $\mathcal{I}_{2}$ transmit data. Each subset can be divided into the left and the right half. The cardinalities for $\mathcal{I}_{0}\sim\mathcal{I}_{2}$ are: 
\begin{math}
\mathrm{Card}\{\mathcal{I}_{0}\}=2,\ \mathrm{Card}\{\mathcal{I}_{1}\}=N_{n},\ \mathrm{Card}\{\mathcal{I}_{2}\}=N_{u}\triangleq N-N_{n}-2  
\end{math}

For the $q$-th block, an alternative sequence can be formulated by taking the conjugation of the right half of $\mathbf{R}_{q}$: 
\begin{equation}
\mathbf{R}_{q}'\triangleq \big[\underbrace{R_{q,0}\ \cdots\ R_{q,N/2-1}}_{N/2}\ \underbrace{R_{q,N/2}^{*}\ \cdots\ R_{q,N-1}^{*}}_{N/2}\big]^{T}\label{equ:Rreverse}
\end{equation}

Now, the \emph{autocorrelation matrix} of $\mathbf{R}_{q}'$ could be written into
\begin{equation}
\boldsymbol\Sigma_{\mathbf{R}_{q}'}=\begin{bmatrix}
\sigma_{W}^{2} & \mathbf{0}_{1\times (N-1)}\\
\mathbf{0}_{(N-1)\times 1} & \mathbf{B}
\end{bmatrix}
\end{equation}
where the sub-matrix $\mathbf{B}$ is shown in \eqref{equ:BSBN} on the next page.  
\begin{figure*}[!t]
\normalsize
\begin{equation}
\mathbf{B}=\underbrace{\begin{bmatrix}
\sigma_{X}^{2}\sigma_{H}^{2}\mathbf{I}_{N_{u}/2} & \mathbf{0}_{N_{u}/2\times (N_{n}+1)} & \left(e^{j\Theta_{q,\boldsymbol\theta}}\sigma_{X}^{2}\sigma_{H}^{2}\right)\mathbf{J}_{N_{u}/2}\\
\mathbf{0}_{(N_{n}+1)\times N_{u}/2} & \mathbf{0}_{N_{n}+1} & \mathbf{0}_{(N_{n}+1)\times N_{u}/2}\\
\left(e^{j\Theta_{q,\boldsymbol\theta}}\sigma_{X}^{2}\sigma_{H}^{2}\right)\mathbf{J}_{N_{u}/2} & \mathbf{0}_{N_{u}/2\times (N_{n}+1)} & \sigma_{X}^{2}\sigma_{H}^{2}\mathbf{I}_{N_{u}/2} 
\end{bmatrix}}_{\mathbf{B}_{s}}+\underbrace{\begin{bmatrix}
\sigma_{W}^{2}\mathbf{I}_{N_{u}/2} & \mathbf{0}_{N_{u}/2\times (N_{n}+1)} & \sigma_{W}^{2}\mathbf{J}_{N_{u}/2}\\
\mathbf{0}_{(N_{n}+1)\times N_{u}/2} & \sigma_{W}^{2}\mathbf{C}_{N_{n}+1} & \mathbf{0}_{(N_{n}+1)\times N_{u}/2}\\
\sigma_{W}^{2}\mathbf{J}_{N_{u}/2} & \mathbf{0}_{N_{u}/2\times (N_{n}+1)} & \sigma_{W}^{2}\mathbf{I}_{N_{u}/2} 
\end{bmatrix}}_{\mathbf{B}_{n}}\label{equ:BSBN}
\end{equation}
\hrulefill
\vspace*{4pt}
\end{figure*}
$\mathbf{C}_{N_{n}+1}$ represents the $(N_{n}+1)\times (N_{n}+1)$ bi-diagonal matrix with unity diagonal and anti-diagonal entries\footnote{For instance, \tiny\begin{math}
	\mathbf{I}_{2}=\begin{bmatrix}
	1 & 0\\
	0 & 1 
	\end{bmatrix},
	\mathbf{J}_{2}=\begin{bmatrix}
	0 & 1\\
	1 & 0 
	\end{bmatrix},
	\mathbf{C}_{3}=\begin{bmatrix}
	1 & 0 & 1\\
        0 & 1 & 0\\
	1 & 0 & 1
	\end{bmatrix}
\end{math}\normalsize}, and $\Theta_{q,\boldsymbol\theta}=\Theta_{q,k,\xi,\eta}+\Theta_{q,N-k,\xi,\eta}=2\pi\xi+\eta\left(\pi N+2\pi N_{g}+2\pi q N_{s}\right)\ \forall k$ where $\boldsymbol\theta=[\xi\ \eta]^{T}$. $\mathbf{B}_{n}$ is closely tied with the noise structure since
\begin{math}
\mathbf{B}_{n}=\left[\boldsymbol\Sigma_{\mathbf{W}}\right]_{1:N-1,1:N-1}+\left[\boldsymbol\Sigma_{\mathbf{W}}^{p}\right]_{1:N-1,1:N-1}
\end{math} where $[\cdot]_{N_{1}:N_{2},N_{3}:N_{4}}$ represents the sub-matrix spanning rows $N_{1}\sim N_{2}$ and columns $N_{3}\sim N_{4}$. Discussions of $\mathbf{B}_{n}$ under various scenarios are given in Section \ref{sec:nongaussian}.  

Arranging observations of $Q$ OFDM blocks into a vector \begin{math}
\mathbf{R}'\triangleq[\mathbf{R}_{0}'\ \mathbf{R}_{1}'\ \mathbf{R}_{2}'\ \cdots\ \mathbf{R}_{Q-1}'] 
\end{math}, after manipulations reported in Appendix \ref{sec:LLF}, the \textit{approximated log-likelihood function} conditioned on $\boldsymbol\theta$ under $\Aone$ and $\Atwo$ is
\begin{align}
\Lambda(\mathbf{R}'|\boldsymbol\theta)&=\mathrm{Const.}+2\sum_{q=0}^{Q-1}\sum_{k\in \mathcal{I}_{2}^{+}}\Re\{\lambda_{1}(q,k)e^{-j\Theta_{q,\boldsymbol\theta}}\}\label{equ:LLF}\\
\lambda_{1}(q,k)&=R_{q,k}R_{q,N-k}
\end{align}
$\mathrm{Const.}$ is irrelevant to estimation and can be neglected. $\Lambda(\mathbf{R}'|\boldsymbol\theta)$ coherently accumulates $\Re\{\lambda_{1}(q,k)e^{-j\Theta_{q,\boldsymbol\theta}}\}$ across all $Q$ OFDM blocks and subcarriers in $\mathcal{I}_{2}$. 
\subsection{Joint Estimation of $\xi$ and $\eta$}
Joint estimation of $\boldsymbol\theta$ using \eqref{equ:LLF} takes the form 
\begin{align}
\widehat{\boldsymbol\theta}&=\argmax_{\boldsymbol\theta}\Lambda(\mathbf{R}'|\boldsymbol\theta)\nonumber\\
&\triangleq\argmax_{\boldsymbol\theta}\bigg(\sum_{q=0}^{Q-1}\sum_{k\in \mathcal{I}_{2}^{+}}\Re\{\lambda_{1}(q,k)e^{-j\Theta_{q,\boldsymbol\theta}}\}\bigg)\label{equ:MLE}
\end{align}
where $\widehat{\boldsymbol\theta}$ is the estimation of $\boldsymbol\theta$. \eqref{equ:MLE} requires a 2-D grid search on all possible $\boldsymbol\theta$ for the global maxima. The complexity involved can be formidably high and cannot be pursued in practice \cite{Morelli}. One seemingly plausible solution is to decouple the original 2-D problem into a 1-D one as in \cite{Morelli}. However, it is not possible since estimations of $\xi$ and $\eta$ are \textit{mutually dependent}: estimating $\xi$ by fixing $\eta$ introduces bias into $\widehat{\xi}$ and vice versa. A five-step method is employed to overcome this issue: 
\begin{enumerate}
\item {\bf Ancillary Phases Estimation:} Obtaining ancillary phases $\widehat{\Theta_{q,\boldsymbol\theta}}, \forall q \in [0,Q-1]$ by either a NDA or a DA estimator. For the $q$-th OFDM block,
\begin{equation}\label{equ:optimalest}
\widehat{\Theta_{q,\boldsymbol\theta}}=\begin{cases}
\arg\bigg\{\sum_{k\in \mathcal{I}_{2}^{+}}\lambda_{1}(q,k)\bigg\},&\mathrm{NDA}\\ 
\arg\bigg\{\sum_{k\in \mathcal{I}_{2}^{+}}\lambda_{1}(q,k)\lambda_{2}(q,k)\bigg\},&\mathrm{DA}
\end{cases}
\end{equation}
where $\lambda_{2}(q,k)$ is introduced to highlight the modulation effect and frequency selectivity as
\begin{equation}
	\lambda_{2}(q,k)=\left[2+\SNR^{-1}(q,k)\right]^{-1}\label{equ:lambda2qk}
\end{equation}
where \begin{math}
\SNR(q,k)=\frac{|X_{q,k}|^{2}|H_{q,k}|^{2}}{\sigma_{W}^{2}} 	
\end{math}. This is explained in more details in Appendix \ref{sec:LLF}. For notational convenience, $\mathcal{A}$ denotes either NDA or DA. The DA suits the closed-loop system with feedback to compute $\lambda_{2}(q,k)$. On the other hand, the NDA, with a similar form to that in \cite{aghajeri_blind_2007}, reduces the complexity without calculating $\lambda_{2}(q,k)$.

\item {\bf Phase Unwrapping:} Since $\Theta_{q+1,\boldsymbol\theta}-\Theta_{q,\boldsymbol\theta}=\mathrm{Const.}$, $\{\Theta_{q,\boldsymbol\theta},\ q=0,1,\cdots,Q-1\}$ lie on the same line. However, $\arg$ in \eqref{equ:optimalest} folds $\widehat{\Theta_{q,\boldsymbol\theta}^{\mathcal{A}}}$ into $[-\pi,+\pi]$ and breaks the line into several discrete segments, known as the phase ambiguity. To deal with this issue, phase unwrapping is used to detect the discontinuities in phases sequentially and restore the actual phases. More specifically, once the criterion
\begin{equation}
|\widehat{\Theta_{q+1,\boldsymbol\theta}^{\mathcal{A}}}-\widehat{\Theta_{q,\boldsymbol\theta}^{\mathcal{A}}}|\geq \varphi\pi \label{equ:phaseunwrapping}
\end{equation}
is satisfied, $\pm 2p\pi$ is added to the phases starting from $\widehat{\Theta_{q+1,\boldsymbol\theta}^{\mathcal{A}}}$, until \eqref{equ:phaseunwrapping} is met again (and $\pm 2(p+1)\pi$ is added afterwards). $\varphi\in[0,2]$ is the tolerance level of phase unwrapping. $p$ denotes the number of times that \eqref{equ:phaseunwrapping} is met before $\widehat{\Theta_{q,\boldsymbol\theta}^{\mathcal{A}}}$. The sign in \eqref{equ:phaseunwrapping} is determined by the direction of phase alternation. 
          
\item {\bf Linearization and Least Square Estimation:} The unwrapped phase is denoted as $\widetilde{\Theta_{q,\boldsymbol\theta}^{\mathcal{A}}}$. They could be combined into the vector $\widetilde{\boldsymbol\Theta_{\mathcal{A}}}$. Now, assuming correct phase unwrapping across all the $Q$ OFDM blocks, $\widetilde{\boldsymbol\Theta_{\mathcal{A}}}$ can be embedded into a linear model shown as    
\begin{equation}\label{equ:basiclinear}
\widetilde{\boldsymbol\Theta_{\mathcal{A}}}=\mathbf{E}\boldsymbol\theta_{\mathcal{A}}+\mathbf{V}
\end{equation}
The $Q\times 2$ observation matrix $\mathbf{E}$ can be expressed by
\begin{equation}
\mathbf{E}=\begin{bmatrix} 2\pi & 2\pi & \cdots & 2\pi & \cdots & 2\pi \\ 
D_{0} & D_{1} & \cdots & D_{q} & \cdots & D_{Q-1}\\
\end{bmatrix}^{T}\label{equ:H}
\end{equation}
where $D_{q}=\pi N+2\pi N_{g}+2\pi qN_{s}$ and $\mathbf{V}$ is the self-noise with mean vector $\boldsymbol\mu_{\mathbf{V}}^{\boldsymbol\theta,\SNR}$ and covariance matrix $\boldsymbol\Sigma_{\mathbf{V}}^{\boldsymbol\theta,\SNR}$, expressed by
\begin{align}
\boldsymbol\mu_{\mathbf{V}}^{\boldsymbol\theta,\SNR}&=\left\{-\frac{\sin(\Theta_{q,\boldsymbol\theta})}{\cos(\Theta_{q,\boldsymbol\theta})+\SNR},\ 0\leq q \leq Q-1\right\}\label{equ:muorg}\\
\boldsymbol\Sigma_{\mathbf{V}}^{\boldsymbol\theta,\SNR}&=\mathrm{diag}\left\{\frac{1-\cos(\Theta_{q,\boldsymbol\theta})}{\Xi N \times \SNR},\ 0\leq q \leq Q-1\right\}\label{equ:vorg}
\end{align}
which are derived under $\Atwo$ and $\Athree$, $\Afour$. $\Athree$, $\Afour$ are shown below: 
\begin{description}
\item[$\Athree$] flat fading ($L=1$) with real channel gain $h_{0}(q)$ with zero mean and variance $\sigma_{H}^{2}$. Thus, in frequency domain, $\E\{H_{q,k}\}=0$ and $\E\{|H_{q,k}|^{2}\}=\sigma_{H}^{2}$.
\item [$\Afour$] Constant modulus modulated signal satisfying $\E\{X_{q,k}\}=0$ and $\E\{|X_{q,k}|^{2}\}=\sigma_{X}^{2}$.
\end{description}
Notice that, $\Athree$ and $\Afour$ differ from $\Aone$: they merely impose the finite first moment and second moment on $X_{q,k}$ and $H_{q,k}$ without the Gaussian assumption on $\mathbf{H}_{q}\mathbf{X}_{q}$. $\Xi=\frac{N_{u}}{2N}$ is the ratio between usable pairwise correlations to the useful $\FFT$ duration, and $\SNR=\nu\frac{\sigma_{X}^{2}\sigma_{H}^{2}}{\sigma_{W}^{2}}$ where $\nu=\frac{N_{u}}{N}$ compensates the loss of discarding unused subcarriers. In lack of $\boldsymbol\theta$, it is impossible to pre-compute $\boldsymbol\mu_{\mathbf{V}}^{\boldsymbol\theta,\SNR}$ and $\boldsymbol\Sigma_{\mathbf{V}}^{\boldsymbol\theta,\SNR}$. An ordinary least square (OLS) estimation yields the initial $\widehat{\boldsymbol\theta_{\mathcal{A}}}$: 
\begin{equation}
\widehat{\boldsymbol\theta_{\mathcal{A}}}=(\mathbf{E}^{T}\mathbf{E})^{-1}\mathbf{E}^{T}\widetilde{\boldsymbol\Theta_{\mathcal{A}}}=\boldsymbol\theta_{\mathcal{A}}+\underline{\mathbf{V}}\label{equ:DAlse}
\end{equation}
where $\underline{\mathbf{V}}=(\mathbf{E}^{T}\mathbf{E})^{-1}\mathbf{E}^{T}\mathbf{V}$ is the $2\times 1$ error vector associated with $\mathbf{V}$ by a linear transformation $(\mathbf{E}^{T}\mathbf{E})^{-1}\mathbf{E}$. 

\item {\bf Bias Erasure:} With $\widehat{\boldsymbol\theta_{\mathcal{A}}}$ and assuming perfect a priori knowledge of $\SNR$, both $\boldsymbol\mu_{\mathbf{V}}^{\boldsymbol\theta,\SNR}$ and $\boldsymbol\Sigma_{\mathbf{V}}^{\boldsymbol\theta,\SNR}$ can be replaced by $\boldsymbol\mu_{\mathbf{V}}^{\widehat{\boldsymbol\theta_{\mathcal{A}}},\SNR}$ and $\boldsymbol\Sigma_{\mathbf{V}}^{\widehat{\boldsymbol\theta_{\mathcal{A}}},\SNR}$. \emph{Bias erasure} (BE) of $\boldsymbol\mu_{\mathbf{V}}^{\widehat{\boldsymbol\theta_{\mathcal{A}}},\SNR}$ in \eqref{equ:basiclinear} gives 
\begin{equation}
\underline{\boldsymbol\Theta_{\mathcal{A}}}=\widetilde{\boldsymbol\Theta_{\mathcal{A}}}-\boldsymbol\mu_{\mathbf{V}}^{\widehat{\boldsymbol\theta_{\mathcal{A}}},\SNR}=\mathbf{E}{\boldsymbol\theta_{\mathcal{A}}}+\mathbf{V}'\label{equ:erasure}
\end{equation}
where \begin{math}
\boldsymbol\mu_{\mathbf{V}'}^{\widehat{\boldsymbol\theta_{\mathcal{A}}},\SNR}=\boldsymbol\mu_{\mathbf{V}}^{{\boldsymbol\theta_{\mathcal{A}}},\SNR}-\boldsymbol\mu_{\mathbf{V}}^{\widehat{\boldsymbol\theta_{\mathcal{A}}},\SNR} 
\end{math} and \begin{math}
\boldsymbol\Sigma_{\mathbf{V}'}^{\widehat{\boldsymbol\theta_{\mathcal{A}}},\SNR}=\boldsymbol\Sigma_{\mathbf{V}}^{\widehat{\boldsymbol\theta_{\mathcal{A}}},\SNR}
\end{math}.

\item {\bf Re-estimation:} Updating the estimation using the weighted least square estimation after bias erasure (WLS--BE) by 
\begin{align}
\underline{\boldsymbol\theta_{\mathcal{A}}}\big|_{\mathrm{WLS-BE}}&=(\mathbf{E}^{T}\left[\boldsymbol\Sigma_{\mathbf{V}'}^{\widehat{\boldsymbol\theta_{\mathcal{A}}},\SNR}\right]^{-1}\mathbf{E})^{-1}\nonumber\\
&\mathbf{E}^{T}\left[\boldsymbol\Sigma_{\mathbf{V}'}^{\widehat{\boldsymbol\theta_{\mathcal{A}}},\SNR}\right]^{-1}\underline{\boldsymbol\Theta_{\mathcal{A}}}\label{equ:thetaerasure}
\end{align}
or OLS after bias erasure (OLS--BE) ignoring the weighting factors
\begin{equation}\label{equ:thetaerasureOLS}
\underline{\boldsymbol\theta_{\mathcal{A}}}\big|_{\mathrm{OLS-BE}}=(\mathbf{E}^{T}\mathbf{E})^{-1}\mathbf{E}^{T}\underline{\boldsymbol\Theta_{\mathcal{A}}}
\end{equation}
The performance loss of \eqref{equ:thetaerasureOLS} comparing with \eqref{equ:thetaerasure} could be neglected, since $\boldsymbol\Sigma_{\mathbf{V}'}^{\widehat{\boldsymbol\theta_{\mathcal{A}}},\SNR}$ approaches a diagonal matrix with nearly identical entries as $\cos(\Theta_{q,\boldsymbol\theta})$ varies slowly block-wise given small $\eta$.
\end{enumerate}

\subsection{Variance and Bias of Estimation}
$\Var\{\widehat{\xi}\}$, $\Var\{\widehat{\eta}\}$, $\mathrm{Bias}\{\widehat{\xi}\}$, and $\mathrm{Bias}\{\widehat{\eta}\}$ before BE with perfect phase unwrapping can be computed straightforwardly from \eqref{equ:basiclinear} under $\Atwo$, $\Athree$, and $\Afour$. The details are shown in Appendix \ref{sec:stat}. Analytical expressions of $\Var\{\widehat{\xi}\}$,$\Var\{\widehat{\eta}\}$,$\mathrm{Bias}\{\widehat{\xi}\}$, and $\mathrm{Bias}\{\widehat{\eta}\}$ are given by \eqref{equ:vareps},\eqref{equ:vareta},\eqref{equ:biaseps},\eqref{equ:biaseta} respectively.  

\begin{figure*}[!t]
\normalsize
\begin{align}
&\Var\{\widehat{\xi}\}=\frac{2g^{2}(2Q+1)(Q+1)+2g(Q+1)(4Q-1)+(2Q+1)(2Q-1)}{4\pi^{2}N(1+g)^{2}Q(Q^2-1)\Xi\mathrm{SNR}}\nonumber\\
&\qquad\quad-\frac{\sum_{q=0}^{Q-1}\cos(\Theta_{q,\boldsymbol\theta})\bigg[(Q-1)\big[1+4g+4Q(1+g)\big]-6q\big[g+Q(1+g)\big]\bigg]^{2}}{4\pi^{2}(1+g)^{2}NQ^{2}(Q+1)^{2}(Q-1)^{2}\Xi\mathrm{SNR}}\label{equ:vareps}\\
&\Var\{\widehat{\eta}\}=\frac{3}{\pi^{2}N^{3}(1+g)^{2}Q(Q^{2}-1)\Xi\mathrm{SNR}}-\frac{\sum_{q=0}^{Q-1}9\cos(\Theta_{q,\boldsymbol\theta})(2q-Q+1)^{2}}{\pi^{2}(1+g)^{2}N^{3}Q^{2}(Q+1)^{2}(Q-1)^{2}\Xi\mathrm{SNR}}\label{equ:vareta}\\
&\mathrm{Bias}\{\widehat{\xi}\}=-\sum_{q=0}^{Q-1}\frac{\sin(\Theta_{q,\boldsymbol\theta})}{\cos(\Theta_{q,\boldsymbol\theta})+\mathrm{SNR}}\bigg[\frac{(Q-1)\big[3+6g+2(1+g)(2Q-1)\big]-6q\big[1+2g+(1+g)(Q-1)\big]}{2\pi(1+g)Q(Q+1)(Q-1)}\bigg]\label{equ:biaseps}\\
&\mathrm{Bias}\{\widehat{\eta}\}=\sum_{q=0}^{Q-1}\frac{\sin(\Theta_{q,\boldsymbol\theta})}{\cos(\Theta_{q,\boldsymbol\theta})+\mathrm{SNR}}\bigg[\frac{3(Q-1)-6q}{\pi(1+g)NQ(Q+1)(Q-1)}\bigg]\label{equ:biaseta}
\end{align}
\hrulefill
\vspace*{4pt}
\end{figure*}

Here, $g=\frac{N_{g}}{N}$ is the ratio between CP and useful $\FFT$ duration with details in Appendix \ref{sec:stat}. The mean square errors ($\mathrm{MSE}$s) are 
\begin{align}
&\MSE\{\widehat{\xi}\}=\Var\{\widehat{\xi}\}+[\mathrm{Bias}\{\widehat{\xi}\}]^{2}\label{equ:MSEXi}\\
&\MSE\{\widehat{\eta}\}=\Var\{\widehat{\eta}\}+[\mathrm{Bias}\{\widehat{\eta}\}]^{2}\label{equ:MSEEta}
\end{align}
With perfect BE, $\boldsymbol\mu_{\mathbf{V}'}^{\widehat{\boldsymbol\theta_{\mathcal{A}}},\SNR}=\mathbf{0}_{Q\times 1}$, and consequently the biases in \eqref{equ:biaseps} and \eqref{equ:biaseta} vanish which makes $\mathrm{MSE}\{\widehat{\xi}\}=\Var\{\widehat{\xi}\}$ and $\mathrm{MSE}\{\widehat{\eta}\}=\Var\{\widehat{\eta}\}$. 

\emph{Remarks:}\\
\indent \emph{(i)}: The first terms of \eqref{equ:vareps} and \eqref{equ:vareta} are written as $\Var\{\widehat{\xi}\}_{1}$ and $\Var\{\widehat{\eta}\}_{1}$. It is easy to prove that \begin{math}
\max\left[\Var\{\widehat{\xi}\}\right]=2\Var\{\widehat{\xi}\}_{1}
\end{math} and \begin{math}
\max\Big[\Var\{\widehat{\eta}\}\Big]=2\Var\{\widehat{\eta}\}_{1}
\end{math}. They provide a good criterion on the choice of $Q$ and $N$. For instance, when $N$ is fixed, to warrant the performance of $\MSE\{\widehat{\xi}\}_{Target}$ and $\MSE\{\widehat{\eta}\}_{Target}$ under a certain $\SNR$ assuming perfect BE, $Q$ should be chosen according to \eqref{equ:Qcriterion} shown on the next page, for $Q\gg 1$ by setting \begin{math}
\MSE\{\widehat{\xi}\}_{Target}=\max\left[\Var\{\widehat{\xi}\}\right]	
\end{math} and \begin{math}
\MSE\{\widehat{\eta}\}_{Target}=\max\Big[\Var\{\widehat{\eta}\}\Big]	
\end{math}; $\Omega$ denotes the $\SNR$ gap considering the possible loss of accuracy of \eqref{equ:vareps} and \eqref{equ:vareta}.   

\begin{figure*}[!t]
\normalsize
\begin{equation}
Q=\max\left[\frac{2}{N}\left(\MSE\{\widehat{\xi}\}_{Target}\pi^{2}\Xi(\SNR +\Omega)\right)^{-1},\frac{\sqrt[3]{6}}{N}\Big(\MSE\{\widehat{\eta}\}_{Target}\pi^{2}(1+g)^{2}\Xi(\SNR+\Omega)\Big)^{-\frac{1}{3}}\right]\label{equ:Qcriterion} 
\end{equation}
\hrulefill
\vspace*{4pt}
\end{figure*}

\indent \emph{(ii)}: Obviously, $\lambda_{2}(q,k)$ in \eqref{equ:lambda2qk} reduces to $\SNR(q,k)$ in low $\SNR$, and becomes $0.5$ in high $\SNR$. Thus, the advantage of DA is more significant in low $\SNR$.        

\begin{figure*}[tb]
\centering
\includegraphics[scale=0.55]{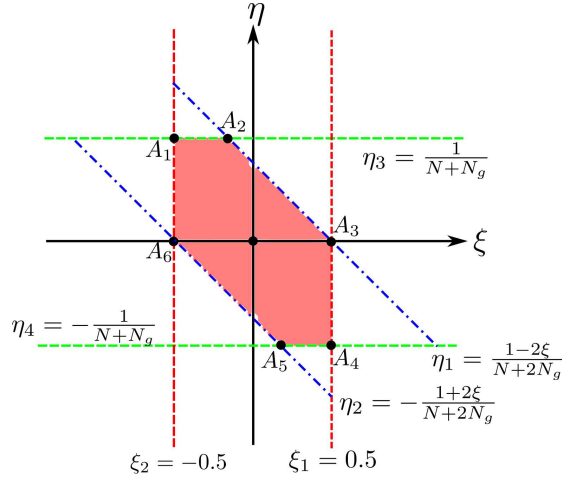}
\caption{The region of $\xi$ and $\eta$ for perfect phase unwrapping in a noiseless environment. The coordinates for $A_{1}\sim A_{6}$ are: $A_{1}:(-0.5,(N+N_{g})^{-1}),\ A_{2}:(-N_{g}\left(2(N+N_{g})\right)^{-1},(N+N_{g})^{-1}),\ A_{3}:(0.5,0),\ A_{4}:(0.5,(N+N_{g})^{-1}),\ A_{5}:(N_{g}\left(2(N+N_{g})\right)^{-1},-(N+N_{g})^{-1}),\ A_{6}:(-0.5,0)$.}
\label{fig:3}
\end{figure*}

\indent \emph{(iii)}: For correct phase unwrapping in absence of noise, two conditions must be satisfied:
\begin{description}
\item[$\mathrm{(C1)}$] $-\pi \leq  2\pi \xi +\pi \eta N + 2\pi N_{g}\eta \leq +\pi$ for correct $\xi$ estimation
\item [$\mathrm{(C2)}$] $-2\pi \leq 2\pi N_{s}\eta \leq 2\pi$ for correct $\eta$ estimation   
\end{description}
Note that, only $\mathrm{Frac}\{\xi\}$ is estimable, since $\mathrm{Int}\{\xi\}$ results in an imperceptible $2\pi \mathrm{Int}\{\xi\}$ rotation in $\Theta_{q,\boldsymbol\theta}$; the only impact of $\mathrm{Int}\{\xi\}$ on $\mathrm{Frac}\{\xi\}$ is an increased $\mathrm{ICI}_{\xi,\eta;q,k}$. Thus, $\xi$ is treated equivalently as $\mathrm{Frac}\{\xi\}$ and bounded in $[-0.5,0.5]$. The \emph{hexagon} denoting the solution area is plotted in Fig.~\ref{fig:3}, with area $S=\frac{3}{2(N+N_{g})}$, which shrinks in presence of noise.
\\
\indent \emph{(iv)}: In practice, $\lambda_{2}(q,k)$ (for the DA scheme) and $\SNR$ (for both DA and NDA schemes) need to be calculated. Firstly, $\lambda_{2}(q,k)$ in \eqref{equ:lambda2qk} is computed as  
\begin{equation}\label{equ:lambda2qkest}
	\widehat{\lambda_{2}(q,k)}=\left[2+\overline{\SNR^{-1}(q,k)}\right]^{-1}
\end{equation}
where \begin{math}
\overline{\SNR^{-1}(q,k)}=\left(\frac{|\overline{X_{q,k}}|^{2}|\widehat{H_{q,k}}|^{2}}{\widehat{\sigma_{W}^{2}}}\right)^{-1}
\end{math}. $\overline{X_{q,k}}$ is either $X_{q,k}$ (for pilots or null subcarriers), or the output of decision device $\widehat{X_{q,k}}$. $\widehat{H_{q,k}}$ stands for the estimated CTF. With \eqref{equ:BSBN}, $\sigma_{W}^{2}$ and $\SNR$ can be estimated by  
\begin{align}
\widehat{\sigma_{W}^{2}}&=\frac{2\sum_{q=0}^{Q-1}\sum_{k \in \mathcal{I}_{1}^{+}}\lambda_{1}(q,k)}{QN_{n}}\label{equ:sigmaWest}\\
\widehat{\SNR}&=\nu\frac{\widehat{\sigma_{X}^{2}\sigma_{H}^{2}}}{\widehat{\sigma_{W}^{2}}}=\nu\left(\frac{\widehat{\sigma_{\mathcal{X}}^{2}}}{\widehat{\sigma_{W}^{2}}}-1\right)\label{equ:SNR}
\end{align}
where 
\begin{equation}
\widehat{\sigma_{\mathcal{X}}^{2}}\triangleq\widehat{\sigma_{X}^{2}\sigma_{H}^{2}}+\widehat{\sigma_{W}^{2}}=\frac{2\sum_{q=0}^{Q-1}\sum_{k \in \mathcal{I}_{2}^{+}}|R_{q,k}|^{2}}{QN_{u}}\label{equ:XHW}
\end{equation}
Now, $\widehat{\SNR}$ is inserted back into \eqref{equ:muorg} and \eqref{equ:vorg} to yield $\boldsymbol\mu_{\mathbf{V}}^{\widehat{\boldsymbol\theta_{\mathcal{A}}},\widehat{\SNR}}$. Replacing $\boldsymbol\mu_{\mathbf{V}}^{\widehat{\boldsymbol\theta_{\mathcal{A}}},\SNR}$ in \eqref{equ:erasure} with $\boldsymbol\mu_{\mathbf{V}}^{\widehat{\boldsymbol\theta_{\mathcal{A}}},\widehat{\SNR}}$ leads to 
\begin{math}
\underline{\boldsymbol\Theta_{\mathcal{A}}}'=\widetilde{\boldsymbol\Theta_{\mathcal{A}}}-\boldsymbol\mu_{\mathbf{V}}^{\widehat{\boldsymbol\theta_{\mathcal{A}}},\widehat{\SNR}}
\end{math}. Using OLS--BE in \eqref{equ:thetaerasureOLS} gives the practical estimation $\underline{\boldsymbol\theta_{\mathcal{A}}}\big|_{\mathrm{PS}}$, termed as the \emph{practical scheme} (PS) given by: 
\begin{equation}\label{equ:PS}
\underline{\boldsymbol\theta_{\mathcal{A}}}\big|_{\mathrm{PS}}=(\mathbf{E}^{T}\mathbf{E})^{-1}\mathbf{E}^{T}\underline{\boldsymbol\Theta_{\mathcal{A}}}'
\end{equation}

\section{Applicability to Different Types of Noise}\label{sec:nongaussian}
\begin{table*}[tb]
\centering
\caption{Statistical information for Class-A and Nakagami-m distribution}
\label{tab:classAandNaka}
\begin{tabular}{cccc}
\hline\hline
Noise & Parameters & PDF (time domain) & PDF (frequency domain)\\ 
\hlinew{1pt}
\multirow{3}*{Class-A} & \multirow{3}*{$A,T$} & $f(w)=\sum_{p=0}^{\infty}\frac{\alpha_{p}}{\sqrt{2\pi}\sigma_{p}}\exp(-\frac{w^{2}}{2\sigma_{p}^{2}})$ & $W=\mathcal{F}\{w\}$\\
& & $\alpha_{p}=e^{-A}\frac{A^{p}}{p!}$ & $W\sim\mathcal{CN}(0,\frac{T+1}{T}\sigma_{g}^{2})$\\
& & $\sigma_{p}^{2}=\sigma_{g}^{2}\frac{p/A+T}{T}$ & \\
\hline
\multirow{4}*{Nakagami-m} & \multirow{4}*{$m,\ \Omega$} & $f(u)=\frac{2m^{m}(u)^{2m-1}}{\Gamma(m)\Omega^{m}}e^{-\frac{mu^{2}}{\Omega}}$ & $W=\mathcal{F}\{\Re\{w\}\}$\\
& & $m=\frac{\Omega^{2}}{\E\{(u^{2}-\Omega)^{2}\}}$ & $W\sim\mathcal{CN}(0,\Omega)$\\
& & $\Omega = \E\{u^{2}\}$ & \\
& & $\Re\{w\}=\sqrt{2}u\cos(\zeta)$,\ $\zeta\sim\mathcal{U}(-\pi,+\pi)$ & \\ 
\hlinew{1pt}
\end{tabular}
\end{table*}
\subsection{Class-A Impulsive Noise}
The PDF for time domain real Class-A noise random variable (r.v.) $w$ is shown in Table~\ref{tab:classAandNaka} \cite{Middleton}. $\mathcal{F}$ denotes the $\FFT$ operation. $T=\sigma_{g}^{2}/\sigma_{i}^{2}$ is the power ratio between the Gaussian component with variance $\sigma_{g}^{2}$ and the additional man-made noise component with variance $\sigma_{i}^{2}$. The Class-A model combines both components and its PDF can be regarded as a weighted sum of infinitely many Gaussian PDFs with an increasing variance $\sigma_{p}^{2}$. Impulsive index $A$ is the product of the average rate of impulsive noise and the mean duration of a typical impulse. The noise is \emph{more impulsive} for a small $A$, whereas for $A \rightarrow \infty$, the PDF becomes a Gaussian distribution \cite{Haring}. After $\FFT$, $W$ in frequency domain approaches Gaussianity, particularly for a large $A$, which is accurate for $p>3$ \cite{berry_understanding_1981}. 

\subsection{Nakagami-m Background Noise}
The envelope PDF of time domain Nakagami-m background noise r.v. $u$ is given in Table~\ref{tab:classAandNaka} \cite{kim_closed-form_2008}, where $m$ is the \textit{fading figure} controlling the severity of the amplitude fading \cite{Beaulieu} and $m\geq 0.5$. $\Gamma(\cdot)$ is the Gamma function, and $\Omega$ the \textit{second moment} of $u$ \cite{Julian}. For the corresponding complex random noise $w$ with an envelope following the Nakagami-m PDF ($|w|=u$), the \emph{axis-PDF} for its real part $\Re\{w\}=u\cos(\zeta)$ is shown in power series in \cite{Meng} and in closed-form in \cite{kim_closed-form_2008}; $\zeta$ is uniformly distributed in $[-\pi,+\pi]$. Yet, the closed-form distribution for the corresponding frequency domain r.v. $W$ is hard to pursue. As a rule of thumb, $W$ can be modeled as a Gaussian r.v. with zero mean and variance $\Omega$, by virtue of the central limit theorem and the uniformly distributed $\zeta$ \cite{Meng}. An additional $\sqrt{2}$ is inserted to compensate the loss of ignoring $\Im\{w\}$. 

\textit{Remarks:}\\
\indent \emph{(i)}: In a strict sense, $m$ differs in different frequency ranges. For high frequencies, $m\approx 1$, while $m<1$ for low frequencies \cite{Meng}. For simplicity, a common $m$ for all subcarriers is assumed.   

\indent \emph{(ii)}: A good tool to evaluate the goodness-of-fit is the $\chi^{2}$-test. In this paper, the D'Agostino's $\chi^{2}$-test is adopted with a significance level $\alpha=0.05$, which is based on the transformation of the third and fourth-order statistics (skewness and kurtosis) on data samples \cite{agostino}. $T$ is set as $\{1,0.1,0.01\}$ with respect to light, moderate, and heavy impulsiveness, while $A\in\{0.01,0.1,1\}$. For the Nakagami-m background noise, $m\in\{0.5, 1.0, 2.0, 10\}$. The kurtosis and skewness for the Gaussian case are $3$ and $0$ respectively. 

Independent and identically distributed (i.i.d) noise samples are generated using the corresponding PDFs. The averaged in-phase p-value (probability that the distribution cannot be rejected as Gaussian) on usable subcarriers with $N=512$ is shown in Table~\ref{tbu:kurtosis} \footnote{For the Nakagami-m background noise, the DC subcarrier could reach a considerably high value which would destroy the Gaussianity. Fortunately, it is usually discarded in most OFDM systems.}. Also, the kurtosis and skewness are tabulated. Similar results can be obtained for the quadrature. For most cases, near-Gaussian performances can be achieved. Thus, $\boldsymbol\Sigma_{\mathbf{W}}$, $\boldsymbol\Sigma_{\mathbf{W}}^{p}$, and $\mathbf{B}_{n}$ in \eqref{equ:BSBN} are invariant. The Gaussianity degrades for $A=0.01$ with $T=0.01$ or $T=0.1$.   

\begin{table*}[tb]
\centering
\caption{Kurtosis, Skewness, and p-value of the D'Agostino's $\chi^{2}$-test for Class-A and Nakagami-m noise ($N=512,\alpha=0.05$)}
\begin{tabular}{|c|c|c|c|c|c|}   
\hlinew{1pt}
Noise & 
\multicolumn{2}{c|}{Parameters} & Kurt.(in-phase) & Skew.(in-phase) & p-value(in-phase) \\
\hlinew{1pt}
\multirow{9}*{Class-A} & \multirow{3}*{$T=0.01$} & $A=0.01$ & $\mathbf{2.61}$ & $\mathbf{9.12\times 10^{-4}}$ & $\mathbf{0.5929}$ \\
\cline{3}
& & $A=0.1$ & $2.94$ & $7.21\times 10^{-4}$ & $0.9532$\\
\cline{3}
& & $A=1$ & $2.98$ & $2.10\times 10^{-3}$ & $0.9521$ \\
\cline{2-3}
& \multirow{3}*{$T=0.1$} & $A=0.01$ & $\mathbf{2.69}$ & $\mathbf{-2.7\times 10^{-3}}$ & $\mathbf{0.7304}$\\
\cline{3}
& & $A=0.1$ & $2.94$ & $-1.3\times 10^{-4}$ & $0.9544$\\
\cline{3}
& & $A=1$  & $2.98$ & $8.75\times 10^{-4}$ & $0.9498$\\
\cline{2-3} 
& \multirow{3}*{$T=1$} & $A=0.01$ & $2.90$ & $6.78\times 10^{-4}$ & $0.9435$\\
\cline{3}
& & $A=0.1$ & $2.98$ & $1.90\times 10^{-3}$ & $0.9502$\\
\cline{3}
& & $A=1$  & $2.98$ & $-8.1\times 10^{-4}$ & $0.9461$\\
\hlinew{1pt}
\multirow{4}*{Nakagami-m} & \multicolumn{2}{c|}{$m=0.5$} & $2.98$ & $-8.3\times 10^{-4}$ & $0.9154$\\
\cline{2-3}
& \multicolumn{2}{c|}{$m=1.0$} & $2.99$ & $-1.4\times 10^{-3}$ & $0.9173$\\
\cline{2-3}
& \multicolumn{2}{c|}{$m=2.0$} & $2.98$ & $-2.4\times 10^{-4}$ & $0.9116$\\
\cline{2-3}
& \multicolumn{2}{c|}{$m=10.0$} & $2.98$ & $-1.1\times 10^{-3}$ & $0.9112$\\
\hlinew{1pt}
\end{tabular}
\label{tbu:kurtosis}
\end{table*}


\subsection{Effect of Temporal Correlation and Cyclo-stationarity}
For colored noise, the zero vectors in $\mathbf{B}_{n}$ are replaced by non-zero values, and both the diagonal and the anti-diagonal elements of $\mathbf{B}_{n}$ are not uniform and this degrades the estimation. The cyclo-stationarity could be analyzed in a similar way. The uniformity of the (anti-)diagonal elements in $\mathbf{B}_{n}$ is \emph{preserved} which entails no performance loss. Meanwhile, non-zero elements appear in the original zero vectors in $\mathbf{B}_{n}$. As the zero vectors in $\mathbf{B}_{n}$ are not involved in the estimation, they would not degrade the performance.

\section{Simulation}\label{sec:sim}
An OFDM system with the following parameters is considered, unless otherwise mentioned:
\begin{enumerate}
\item \textbf{Sampling interval}: $T_{sam}=10ns$ (sampling frequency $100$ MHz)
\item \textbf{Number of subcarriers}: $N=512$
\item \textbf{Guard interval}: $N_{g}=64$
\item \textbf{OFDM blocks}: $Q=10$
\item \textbf{Null subcarriers}: $N_{n}=64$ 
\item \textbf{Modulation}: 16-$\PSK$ ($M=16, \sigma_{X}^{2}=1$)
\item \textbf{Channel}: The real channel taps are generated from the model proposed in \cite{Tonello}. Channel classes $1\sim 9$ with different channel capacities are selected with respect to the probability profile \{0.0349,\ 0.1678,\ 0.1818,\ 0.1188,\ 0.1188,\ 0.1258,\ 0.0979,\ 0.0769,\ 0.0769\}. The number of paths $L$ follows a Poisson distribution. Also, the channel length is tailored to satisfy $L\leq \frac{N_{g}}{2}=32$. Other parameters are chosen according to \cite{Tonello}.
\item \textbf{Impulsive parameters}: $A,T\in \{0.01,0.1,1\}\times \{0.01,0.1,1\}$
\item \textbf{Nakagami-m fading figure}: $m\in \{0.5,1.0,2.0,10.0\}$
\item \textbf{SPO}: $\xi=0.1$
\item \textbf{SCO}: $\eta=1\times 10^{-5}$
\item \textbf{Phase unwrapping tolerance level}: $\varphi=1$ in \eqref{equ:phaseunwrapping}
\end{enumerate}

We generate i.i.d impulsive Class-A noise samples using the toolbox in \cite{Evans}, while i.i.d Nakagami-m background noise samples by the square root of Gamma distributed r.v.. $\mathrm{MSE}\{\widehat{\xi}\}$ and $\mathrm{MSE}\{\widehat{\eta}\}$ are chosen as the criteria given by $\mathrm{MSE}\{\widehat{\xi}\}=\E\{|\widehat{\xi}-\xi|^{2}\}$ and $\mathrm{MSE}\{\widehat{\eta}\}=\E\{|\widehat{\eta}-\eta|^{2}\}$. The achievable variances after BE are plotted as $\Var\{\widehat{\xi}\}$ and $\Var\{\widehat{\eta}\}$ using the first terms of \eqref{equ:vareps} and \eqref{equ:vareta} as benchmarks.

As far as the colored noise, the $1/f^{\beta}$ model is used with $\beta\in \{0.72,0.337\}$ for generation of correlated noise samples \cite{Esmailian}. The condition $\mathrm{Tr}\{\mathbf{B}_{n}\}=(N-1)\sigma_{W}^{2}$ is imposed to keep the total noise power invariant.

For the cyclo-stationary noise, similar to the noise model proposed in \cite{Katayama}, the instantaneous variance is set as 
\begin{equation}
	\E\{w(t)\}\triangleq\sigma_{W}^{2}(t)=A^{2}(t)\sin^{2}\left(\frac{2\pi t}{T_{AC}}\right)
\end{equation}
where $T_{AC}$ is the main voltage frequency ($60\ \mathrm{Hz}$ here). $A_{q}$ is chosen according to \eqref{equ:A} to make the average energy in the $q$-th block $[(n_{0}+(q-1)N)T_{sam},(n_{0}+qN-1)T_{sam}]$ invariant:
\begin{align}
A_{q}&=A(t)\big|_{t=(n_{0}+(q-1)N)T_{sam},\cdots,(n_{0}+qN-1)T_{sam}}\nonumber\\
&=\left[\frac{N\sigma_{W}^{2}}{\sum_{n=n_{0}+(q-1)N}^{n_{0}+qN-1}\sin^{2}\left(\frac{2\pi nT_{sam}}{T_{AC}}\right)}\right]^{1/2}\label{equ:A}
\end{align}

\begin{figure}[tbp]
\centering
\includegraphics[width=20pc]{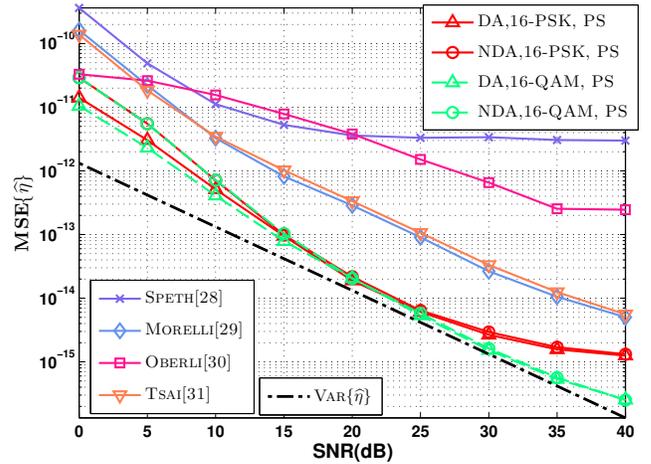}
\caption{Comparison of $\mathrm{MSE}\{\widehat{\eta}\}$ with PS and other schemes.}
\label{fig:mseeta}
\end{figure}
\begin{figure}[tbp]
\centering
\includegraphics[width=20pc]{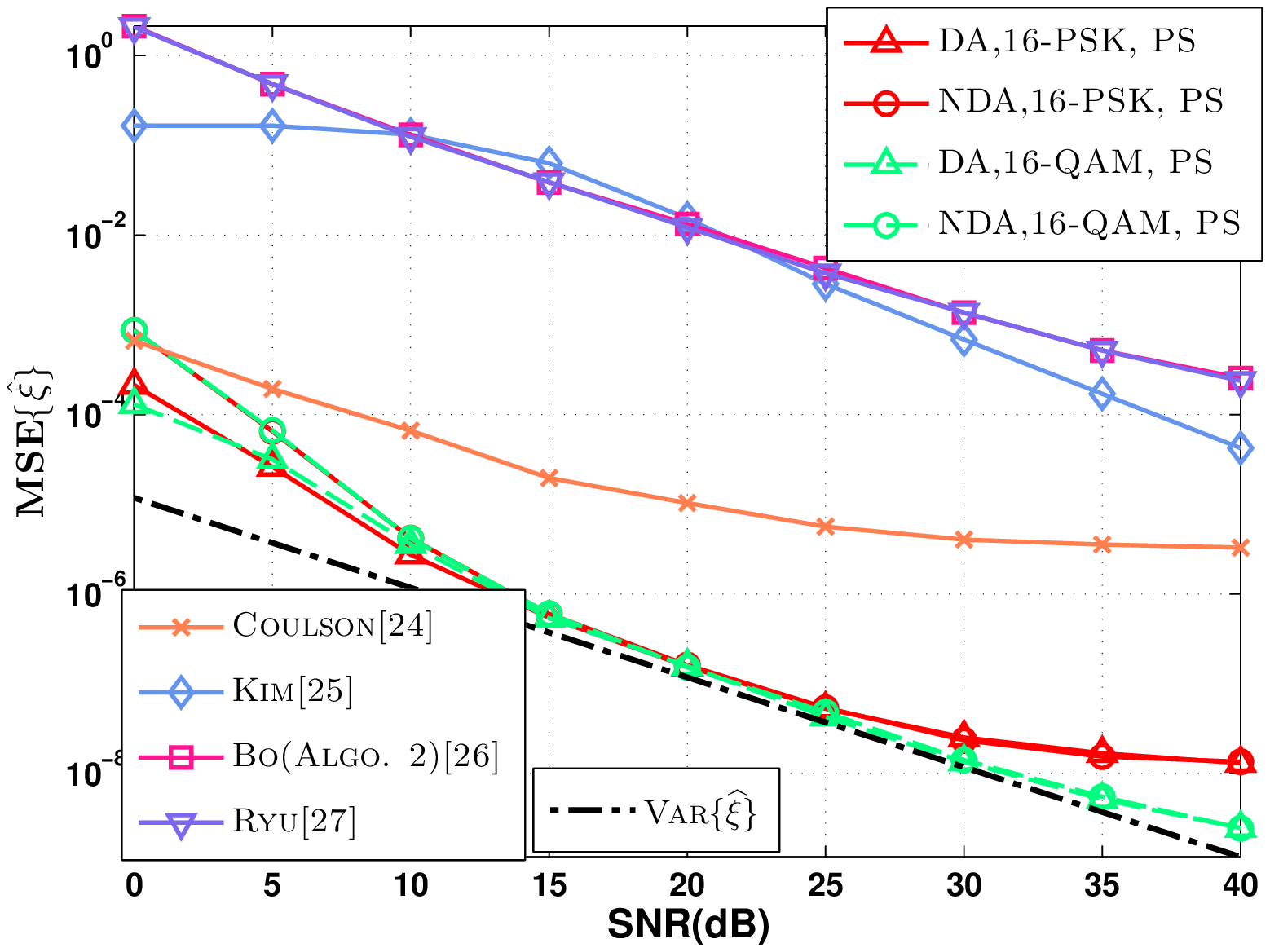}
\caption{Comparison of $\mathrm{MSE}\{\widehat{\xi}\}$ with PS and other schemes.}
\label{fig:mseeps}
\end{figure}
\begin{figure}[tb]
\centering
\includegraphics[width=20pc]{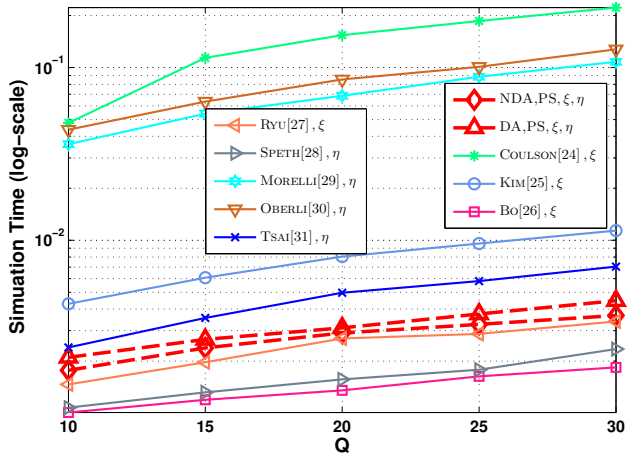}
\caption{Simulation time under different OFDM block number $Q$.}
\label{fig:timing}
\end{figure}
\subsection{Comparison of $\xi$ and $\eta$ estimation performance using PS with other schemes}\label{sec:simA} 
Fig.~\ref{fig:mseeta} and Fig.~\ref{fig:mseeps} demonstrate $\mathrm{MSE}\{\widehat{\eta}\}$ and $\mathrm{MSE}\{\widehat{\xi}\}$ performances of several schemes versus $\SNR$ in additive white Gaussian noise ($\AWGN$). All available subcarriers are treated as pilots ($N_{p}=N-N_{n}=448$). Perfect knowledge of CSI is assumed for \cite{Coulson,DYK,Bo,Ryu,Morelli,OE,JLS} and DA--PS, while unknown for NDA--PS and \cite{SpethII}. For Oberli's estimator \cite{OE}, $\xi=0$ or otherwise $\widehat{\eta}$ would be biased. Likewise, $\eta=0$ for \cite{Coulson,DYK,Bo,Ryu}. RCE is selected from \cite{Morelli}, and the \emph{Algorithm 2} out of the four schemes in \cite{Bo}. Note that, \cite{Bo} and \cite{Ryu} are originally designed for estimation of both $\mathrm{Int}\{\xi\}$ and $\mathrm{Frac}\{\xi\}$ in a multipath channel without knowledge of CSI. However, simulations indicate that CSI should be obtained to guarantee the accuracy of $\mathrm{Frac}\{\xi\}$.

PS in \eqref{equ:PS} is simulated under: $\Sone$ DA, 16-$\PSK$; $\Stwo$ NDA, 16-$\PSK$; $\Sthree$ DA, 16-$\QAM$; $\Sfour$ NDA, 16-$\QAM$. For the performance of $\widehat{\eta}$, $\Sone$ slightly outperforms $\Stwo$ when $\SNR\leq 20$ dB. Advantage of DA is not significant for moderate $\SNR$ since $\lambda_{2}(q,k)\approx \mathrm{Const.}$. The performance gap is more significant in low $\SNR$. For $\Sthree$ and $\Sfour$, the performance gap is enlarged due to the modulation effect of $\QAM$. In the ICI-dominant region, $\Sthree$ and $\Sfour$ outperform $\Sone$ and $\Stwo$.

Similar conclusions can be drawn on $\widehat{\xi}$ except that $\Sone$ slightly outperforms $\Sthree$ when $\SNR\in[4,15]$ dB. According to the later analysis, the OLS using DA outperforms NDA in low to moderate $\SNR$. However, $\boldsymbol\mu_{\mathcal{A}}^{\boldsymbol\theta,\SNR}$, derived under $\Afour$, departs from the case of $\QAM$. This is more significant for $\xi$ according to \eqref{equ:biaseps} and \eqref{equ:biaseta}. Therefore, an inaccurate OLS might in reverse compensate the BE inaccuracy and leads to better performance after BE. 

As observed in Fig.~\ref{fig:mseeta} and Fig.~\ref{fig:mseeps}, both DA--PS and NDA--PS outperform \cite{SpethII,Morelli,JLS,OE,Coulson,DYK,Bo,Ryu}. It is even more remarkable considering that the pilot-assisted schemes exploit all usable subcarriers and perfect CSI. 

The averaged simulation times of DA--PS, NDA--PS, and \cite{Coulson,DYK,Bo,Ryu,Morelli,OE,JLS,SpethII} are plotted in Fig.~\ref{fig:timing}. The complexity of the proposed PS schemes are moderate among all the schemes. 

\begin{figure}[tbp]
\centering
\includegraphics[width=20pc]{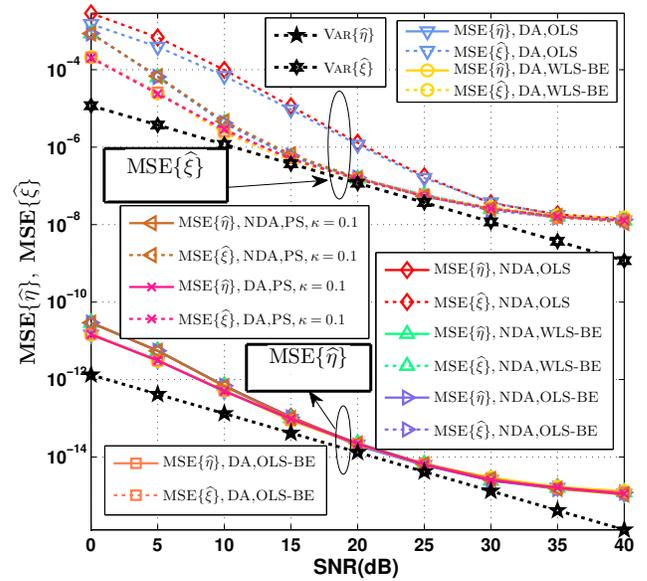}
\caption{Different variants of $\mathrm{MSE}\{\widehat{\eta}\}$ and $\mathrm{MSE}\{\widehat{\xi}\}$ estimation including OLS, WLS--BE, OLS--BE, and PS with DA and NDA.}
\label{fig:comp}
\end{figure}
\subsection{Comparison of the proposed DA/NDA variants} 
Fig.~\ref{fig:comp} compares $\mathrm{MSE}$ performances of OLS \eqref{equ:DAlse}, WLS--BE \eqref{equ:thetaerasure}, OLS--BE \eqref{equ:thetaerasureOLS}, PS \eqref{equ:PS} using either DA or NDA. For the DA--PS scheme, to simulate imperfect channel estimation and therefore imperfect $\lambda_{2}(q,k)$, a $N\times 1$ complex Gaussian random vector is introduced which satisfies HSP. This vector, denoted with $\mathbf{J}_{q}\triangleq[J_{q,0}\ J_{q,1}\ \cdots\ J_{q,N-1}]^{T}$, has zero mean and variance $\frac{\sum_{k=0}^{N-1}|H_{q,k}|^{2}}{N}$ for each OFDM block, independent of $\mathbf{\dot{H}}_{q}$. The imperfect CTF, denoted as $\widetilde{\mathbf{H}_{q}}$, is expressed by \cite{Kyung}
\begin{equation}\label{equ:J}
	\widetilde{\mathbf{H}_{q}}=\sqrt{1-\kappa^{2}}{\mathbf{\dot{H}}_{q}}+\kappa\mathbf{J}_{q}
\end{equation}
where $\kappa$ represents the channel estimation accuracy and is set at $0.1$ here. Before BE, DA--OLS outperforms NDA--OLS in low $\SNR$. BE is important for $\widehat{\xi}$, while $\widehat{\eta}$ does not improve after BE mainly due to imperfect phase unwrapping not considered in derivations. The performances of WLS--BE, OLS--BE and PS are similar. DA outperforms NDA in all situations for low $\SNR$. $\kappa$ does not result in any degradation, since $\lambda_{2}(q,k)$ is not distorted under \eqref{equ:J}.  
\begin{figure}[tbp]
\centering
\includegraphics[width=20pc]{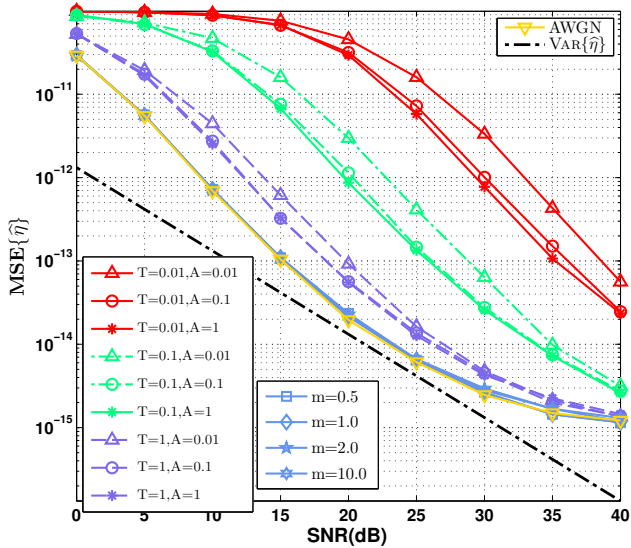}
\caption{$\mathrm{MSE}\{\widehat{\eta}\}$ under Class-A and Nakagami-m background noise using NDA and PS.}
\label{fig:alphagamma}
\end{figure}
\begin{figure}[tbp]
\centering
\includegraphics[width=20pc]{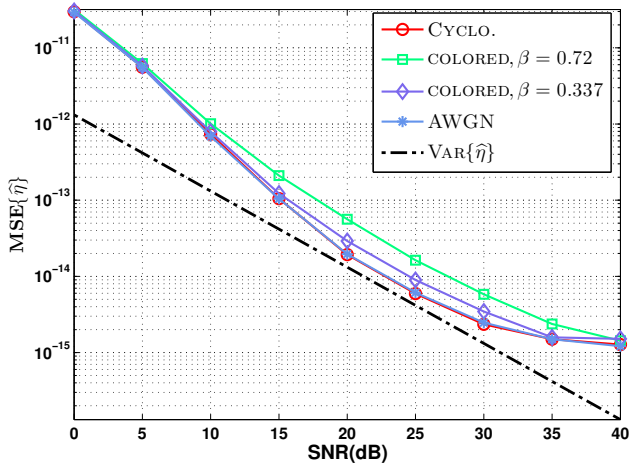}
\caption{$\mathrm{MSE}\{\widehat{\eta}\}$ under cyclo-stationarity and temporal correlation using NDA and PS.}
\label{fig:corNBI}
\end{figure}
\subsection{$\mathrm{MSE}\{\widehat{\eta}\}$ in Different Noise Models}
Fig.~\ref{fig:alphagamma} shows $\mathrm{MSE}\{\widehat{\eta}\}$ under Class-A impulsiveness and Nakagami-m background noise using NDA--PS; in the former case, $\SNR$ is defined as the background $\SNR$ as $\nu\sigma_{X}^{2}\sigma_{H}^{2}/\sigma_{g}^{2}$ to stress the effect of $T$. 

For the Class-A case, a wide margin exists between $\mathrm{MSE}\{\widehat{\eta}\}$ with different $T$. The performance degradation is nearly $7.5$ dB when $T=1\Rightarrow T=0.1$, and $10$ dB when $T=0.1\Rightarrow T=0.01$. The gap between $\AWGN$ and $T=1,A=1$ is $3$ dB. $A=0.01$ degrades the performances for $T=0.01/0.1$ due to the reduction of Gaussianity shown in Table \ref{tbu:kurtosis}, reaching $3.4$ dB and $3.3$ dB for $T=0.1$ and $T=0.01$ respectively; a simple remedy is to increase the $\FFT$ size $N$. The losses for $A=0.1$ and $A=0.01,T=1$ stem from the increased variance of the numerical result $\Var\{W\}=\frac{T+1}{T}\sigma_{g}^{2}$ given a small $A$. In other words, the numerical variance in the Class-A case in Table~\ref{tbu:kurtosis} converges only \emph{in the mean sense} for finite size simulations.  

As far as the Nakagami-m background noise, performances under different $m$ are indistinguishable and approach the performance under $\AWGN$, since the noise variance is determined by $\Omega$ and irrelevant to $m$.

Fig.~\ref{fig:corNBI} shows $\mathrm{MSE}\{\widehat{\eta}\}$ under cyclo-stationarity and temporal correlation using NDA--PS. The cyclo-stationarity entails no performance loss, since the uniformity of the (anti-)diagonal entries of $\mathbf{B}_{n}$ is preserved as mentioned in Section \ref{sec:nongaussian}. On the other hand, the temporal correlation degrades the performances, which is more pronounced for $\beta=0.72$.      

The same conclusions can be made on DA and $\mathrm{MSE}\{\widehat{\xi}\}$ performances which are omitted here due to space limitation.     

\begin{figure}[tbp]
\centering
\includegraphics[width=20pc]{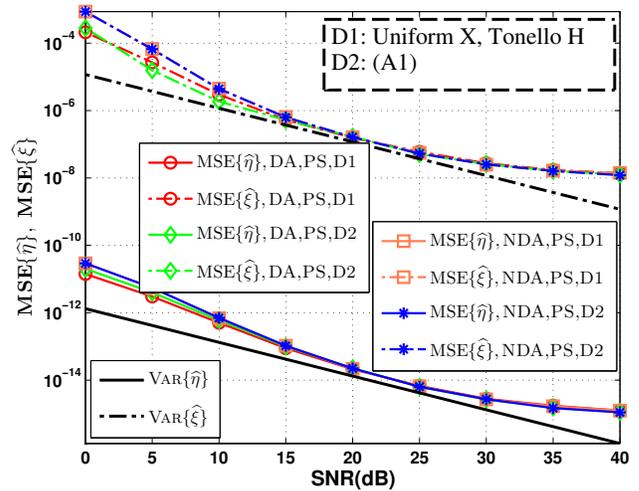}
\caption{Comparison of $\mathrm{MSE}\{\widehat{\eta}\}$ and $\mathrm{MSE}\{\widehat{\xi}\}$ under two statistical assumptions.}
\label{fig:AWGNActual}
\end{figure}
\subsection{$\mathrm{MSE}\{\widehat{\eta}\}$ and $\mathrm{MSE}\{\widehat{\xi}\}$ With Different Statistical Models of Signal and Channel}
Fig.~\ref{fig:AWGNActual} compares $\mathrm{MSE}\{\widehat{\eta}\}$ and $\mathrm{MSE}\{\widehat{\xi}\}$ with PS using DA and NDA under \begin{description}
\item[$\Done$:] Uniform distributed $\mathbf{X}_{q}$ and $\mathbf{\dot{H}}_{q}$ in \cite{Tonello} (Kurtosis: $8.8120$) 
\item[$\Dtwo$:] Gaussian distributed $\mathbf{H}_{q}\mathbf{X}_{q}$ under $\Aone$ (Kurtosis: $2.9741$)
\end{description}
For $\Dtwo$, samples of $\mathbf{H}_{q}\mathbf{X}_{q}$ are generated directly from $\mathcal{CN}(\mathbf{0}_{N\times 1},\sigma_{X}^{2}\sigma_{H}^{2}\mathbf{I}_{N})$. Interestingly, $\MSE\{\widehat{\eta}\}$ with DA under $\Done$ slightly outperforms that under $\Dtwo$, while $\Dtwo$ slightly outperforms $\Done$ for $\SNR\in[2,15]$ dB in terms of $\MSE\{\widehat{\xi}\}$. The kurtosis of $\Done$ is significantly higher than that of $\Dtwo$, signifying a sharper peak and longer tail distribution. Thus, the probability for a considerably high amplitude $\mathbf{H}_{q}\mathbf{X}_{q}$ under $\Done$ outnumbers that of $\Dtwo$, which enhances the profit of $\lambda_{2}(q,k)$ using DA. For $\MSE\{\widehat{\xi}\}$, as stated in \ref{sec:simA}, $\Dtwo$ gains from the reverse compensation. The same performances can be obtained for NDA under $\Done$ and $\Dtwo$. The small performance gap implies the robustness of the proposed technique against violation of $\Aone$ and thus the joint Gaussian assumption of $\mathbf{R}_{q}$. 

\section{Concluding remarks}\label{sec:conclusion}
A novel five-step joint timing acquisition technique is proposed to estimate SPO and SCO. An approximation form of the log-likelihood function is derived under the pairwise correlation and Gaussian assumption on the received data in frequency domain. By using the intra-block phase rotations, the proposed technique achieves significant performance improvements over conventional pilot-assisted estimators relying on inter-block phase rotations. Analytical expressions for the variances and biases are derived. Effects of several noise models including non-Gaussianity, cyclo-stationarity, temporal correlation are discussed. The robustness of the proposed technique against violation of the joint Gaussian assumption is verified by simulations.      

\section{Acknowledgement}
This work is supported by National Science and Technology Major Project of China (Grant No. 2011ZX03003-003-03). The authors would like to thank the anonymous reviewers and the editors who help to improve the quality of this paper. The authors also want to express their gratitudes to the members of WiNLAB and BMCLAB of National Central University for their help.  

\renewcommand{\theequation}{\thesection.\arabic{equation}}
\appendices
\section{Derivation of the approximated log-likelihood function in \eqref{equ:LLF} under $\Aone$ and $\Atwo$}\label{sec:LLF}
Using \eqref{equ:Rreverse} under $\Aone$ and $\Atwo$ in Section \ref{sec:signalmodel}, $\mathbf{R}_{q}$ is a joint Gaussian vector. Similar to \cite{VDB}, the log-likelihood function is derived as 
\begin{align}
\Lambda\left(\mathbf{R}'|\boldsymbol\theta\right)&=\log\bigg(\prod_{q=0}^{Q-1}\prod_{k \in \mathcal{I}_{2}^{+}}f(R_{q,k}',R_{q,N-k}')\bigg)\label{equ:Lambdafunc}\\
f(R_{q,k}',R_{q,N-k}')&=\frac{1}{\pi^{2}\det[\boldsymbol\Sigma_{\mathbf{G}}]}\exp\left(-\mathbf{G}[\boldsymbol\Sigma_{\mathbf{G}}]^{-1}\mathbf{G}^{H}\right)\label{equ:Ffunction}
\end{align}
where $\mathbf{G}\triangleq[R_{q,k}',R_{q,N-k}']$. $\boldsymbol\Sigma_{\mathbf{G}}=\E\{\mathbf{G}^{H}\mathbf{G}\}$ is given by 
\begin{equation}
\boldsymbol\Sigma_{\mathbf{G}}=\begin{bmatrix}
\sigma_{X}^{2}\sigma_{H}^{2}+\sigma_{W}^{2} & e^{j\Theta_{q,\boldsymbol\theta}}\sigma_{X}^{2}\sigma_{H}^{2}+\gamma\sigma_{W}^{2}\\
e^{-j\Theta_{q,\boldsymbol\theta}}\sigma_{X}^{2}\sigma_{H}^{2}+\gamma\sigma_{W}^{2} & \sigma_{X}^{2}\sigma_{H}^{2}+\sigma_{W}^{2}\\ 
\end{bmatrix}
\end{equation}
For convenience, $\gamma$ is introduced which is $1$ for the complete-form log-likelihood function and $0$ for the approximated one. The generalized expression for \eqref{equ:Ffunction} is shown in \eqref{equ:fcomplete}. 
\begin{figure*}[!t]
\normalsize
\begin{equation}
f(R_{q,k}',R_{q,N-k}')=\frac{\exp\left(\frac{-(\sigma_{X}^{2}\sigma_{H}^{2}+\sigma_{W}^{2})(|R_{q,k}|^{2}+|R_{q,N-k}|^{2})+2\Re\{R_{q,k}R_{q,N-k}e^{-j\Theta_{q,\boldsymbol\theta}}\sigma_{X}^{2}\sigma_{H}^{2}\}}{2\sigma_{X}^{2}\sigma_{H}^{2}\sigma_{W}^{2}(1-\gamma\cos(\Theta_{q,\boldsymbol\theta}))+\sigma_{W}^{4}(1-\gamma^{2})}\right)}{\pi^{2}[2\sigma_{X}^{2}\sigma_{H}^{2}\sigma_{W}^{2}(1-\gamma\cos(\Theta_{q,\boldsymbol\theta}))+\sigma_{W}^{4}(1-\gamma^{2})]}\label{equ:fcomplete}
\end{equation}
\hrulefill
\vspace*{4pt}
\end{figure*}
For the complete-form log-likelihood function with $\gamma=1$, it is considerably hard to find a closed-form optimal solution to $\Theta_{q,\boldsymbol\theta}$ due to the presence of $\left[1-\cos(\Theta_{q,\boldsymbol\theta})\right]$. Thus, $\gamma$ is set at $0$ which gives \eqref{equ:LLF}. To fill the gap between the complete and the approximated form of the log-likelihood function, $\sigma_{X}^{2}$ and $\sigma_{H}^{2}$ are replaced by $|X_{q,k}|^{2}$ and $|H_{q,k}|^{2}$ respectively to take advantage of the modulation effect and frequency selectivity, which gives $\lambda_{2}(q,k)$ in \eqref{equ:MLE}.  

\section{Derivation of Variances and Biases in \eqref{equ:vareps}, \eqref{equ:vareta}, \eqref{equ:biaseps}, \eqref{equ:biaseta} under $\Atwo$, $\Athree$, and $\Afour$}\label{sec:stat}
\subsection{Bias Performance}
For DA, subtracting the left hand side of \eqref{equ:optimalest} by the true value of $\Theta_{q,\boldsymbol\theta}$ yields
\begin{equation}
	\Delta\Theta_{q,\boldsymbol\theta}=\widehat{\Theta_{q,\boldsymbol\theta}^{\mathrm{DA}}}-\Theta_{q,\boldsymbol\theta} = \arg\left(\sum_{k\in \mathcal{I}_{2}^{+}}\frac{\lambda_{1}(q,k)\lambda_{2}(q,k)}{e^{j\Theta_{q,\boldsymbol\theta}}} \right) 
\end{equation}
Suppose that $\widehat{\Theta_{q,\boldsymbol\theta}^{\mathrm{DA}}}$ is in vicinity of $\Theta_{q,\boldsymbol\theta}$, $\tan(\Delta\Theta_{q,\boldsymbol\theta})$ can be approximated as 
\begin{equation}\label{equ:vicinity}
\tan(\Delta\Theta_{q,\boldsymbol\theta})=\frac{\Im\left[\Delta\Theta_{q,\boldsymbol\theta}\right]}{\Re\left[\Delta\Theta_{q,\boldsymbol\theta}\right]} \approx \Delta\Theta_{q,\boldsymbol\theta}
\end{equation}
since $\Delta\Theta_{q,\boldsymbol\theta}$ is sufficiently small. Expectation of \begin{math}\Delta\Theta_{q,\boldsymbol\theta}\end{math} in \eqref{equ:vicinity} is approximated by 
\begin{equation}\label{equ:DeltaFrac}
\E\left\{\Delta\Theta_{q,\boldsymbol\theta}\right\} \approx \E \left\{\frac{\Im\left[\Delta\Theta_{q,\boldsymbol\theta}\right]}{\Re\left[\Delta\Theta_{q,\boldsymbol\theta}\right]}\right\} \approx \frac{\E\left\{\Im\left[\Delta\Theta_{q,\boldsymbol\theta}\right]\right\}}{\E\left\{\Re\left[\Delta\Theta_{q,\boldsymbol\theta}\right]\right\}}
\end{equation}
if \begin{math}
\E\left\{\Re\left[\Delta\Theta_{q,\boldsymbol\theta}\right]\right\} \gg \sqrt{\Var \left\{\Re\left[\Delta\Theta_{q,\boldsymbol\theta}\right]\right\}}
\end{math}. Under $\Atwo$, write 
\begin{align}
\E\left\{\Im\left[\Delta\Theta_{q,\boldsymbol\theta}\right]\right\}&=-\sum_{k \in \mathcal{I}_{2}^{+}}\sin(\Theta_{q,\boldsymbol\theta})\lambda_{2}(q,k)\label{equ:E1}\\
\E\left\{\Re\left[\Delta\Theta_{q,\boldsymbol\theta}\right]\right\}&=\sum_{k \in \mathcal{I}_{2}^{+}}\big[\cos(\Theta_{q,\boldsymbol\theta})+\SNR(q,k)\big]\lambda_{2}(q,k)\label{equ:E2}
\end{align}

Under $\Athree$ and $\Afour$, $\E\left\{\Delta\Theta_{q,\boldsymbol\theta}\right\}$ takes the form 
\begin{equation}\label{equ:EoverallSimp}
\E\left\{\Delta\Theta_{q,\boldsymbol\theta}\right\}\approx-\frac{\sin(\Theta_{q,\boldsymbol\theta})}{\cos(\Theta_{q,\boldsymbol\theta})+\SNR}
\end{equation}
where $\SNR=\SNR(q,k)$ and $\lambda_{2}(q,k)=\left(2+\SNR^{-1}\right)=\mathrm{Const.}$. Assuming perfect phase unwrapping, \eqref{equ:DAlse} gives \begin{math}
\mathrm{Bias}\{\widehat{\boldsymbol\theta}_{\mathrm{DA}}\}=\mathrm{Bias}\{\underline{\mathbf{V}}\}
\end{math}. The $2\times 1$ bias vector $\mathrm{Bias}\{\underline{\mathbf{V}}\}$ is associated with the $Q\times 1$ bias vector $\mathrm{Bias}\{\mathbf{V}\}$ by  
\begin{equation}
\mathrm{Bias}\{\underline{\mathbf{V}}\}=(\mathbf{E}^{T}\mathbf{E})^{-1}\mathbf{E}^{T}\mathrm{Bias}\{\mathbf{V}\}
\end{equation}
where \begin{math}
\mathrm{Bias}\{\mathbf{V}\}=[-\frac{\sin(\Theta_{0,\boldsymbol\theta})}{\cos(\Theta_{0,\boldsymbol\theta})+\SNR}\quad -\frac{\sin(\Theta_{1,\boldsymbol\theta})}{\cos(\Theta_{1,\boldsymbol\theta})+\SNR}\quad \cdots\quad -\frac{\sin(\Theta_{Q-1,\boldsymbol\theta})}{\cos(\Theta_{Q-1,\boldsymbol\theta})+\SNR}]^{T}
\end{math}. Standard calculations lead to \eqref{equ:biaseps} and \eqref{equ:biaseta}.
\subsection{Variance Performance}
$\Var\left\{\Delta\Theta_{q,\boldsymbol\theta}\right\}$ can be formulated into 
\begin{equation}\label{equ:Varequ}
\Var\left\{\Delta\Theta_{q,\boldsymbol\theta}\right\} \approx \frac{\Var\{\Im\left[\Delta\Theta_{q,\boldsymbol\theta}\right]\}}{\left(\E\{\Re\left[\Delta\Theta_{q,\boldsymbol\theta}\right]\}\right)^{2}} 
\end{equation}
under \begin{math}
\E\left\{\Re\left[\Delta\Theta_{q,\boldsymbol\theta}\right]\right\} \gg \sqrt{\Var\left\{\Re\left[\Delta\Theta_{q,\boldsymbol\theta}\right]\right\}}
\end{math}. Again, assuming $\Atwo$, $\left(\E\{\Re\left[\Delta\Theta_{q,\boldsymbol\theta}\right]\}\right)^{2}$ can be written into \eqref{equ:EREAL}. 
\begin{figure*}[!t]
\normalsize
\begin{align}
&\left(\E\{\Re\left[\Delta\Theta_{q,\boldsymbol\theta}\right]\}\right)^{2}=\left[\sum_{{k\in \mathcal{I}_{2}^{+}}}\left(|X_{q,k}|^{2}|H_{q,k}|^{2}+\sigma_{W}^{2}\cos(\Theta_{q,\boldsymbol\theta})\right)\lambda_{2}(q,k)\right]^{2}\label{equ:EREAL}\\
&(\Im\left[\Delta\Theta_{q,\boldsymbol\theta}\right])^{2}=\sum_{{k_{1}\in \mathcal{I}_{2}^{+}}}\sum_{{k_{2}\in \mathcal{I}_{2}^{+}}} \bigg(\underbrace{|W_{q,k_{1}}|^{2}|W_{q,k_{2}}|^{2}\sin^{2}(\Theta_{q,\boldsymbol\theta})}_{Y_{1}}-\underbrace{|W_{q,k_{1}}|^{2}\sin(\Theta_{q,\boldsymbol\theta})\Im\{\lambda_{3}(q,k_{2})\}}_{Y_{2}}\nonumber \\
&-\underbrace{|W_{q,k_{2}}|^{2}\sin(\Theta_{q,\boldsymbol\theta})\Im\{\lambda_{3}(q,k_{1})\}}_{Y_{3}} + \underbrace{\Im\{\lambda_{3}(q,k_{1})\}\Im\{\lambda_{3}(q,k_{2})\}}_{Y_{4}} \bigg) \times \underbrace{\lambda_{2}(q,k_{1})\lambda_{2}(q,k_{2})}_{Y_{5}}\label{equ:expandingIm}\\
&\lambda_{3}(q,k)=e^{j\Theta_{q,k-N,\xi,\eta}}X_{q,k}H_{q,k}W_{q,N-k}+e^{j\Theta_{q,-k,\xi,\eta}}X_{q,N-k}H_{q,N-k}W_{q,k}\label{equ:lambda3}\\
&\E\{\sum_{{k_{1} \in \mathcal{I}_{2}^{+}}}\sum_{{k_{2}\in \mathcal{I}_{2}^{+}}} Y_{1}\times Y_{5}\}=\sin^{2}(\MyTheta)\sigma_{W}^{4}\sum_{{k_{1} \in \mathcal{I}_{2}^{+}}}\sum_{{k_{2} \in \mathcal{I}_{2}^{+}}}\lambda_{2}(q,k_{1})\lambda_{2}(q,k_{2})\label{equ:Y1Y5}\\ 
&\E \{\sum_{{k_{1} \in \mathcal{I}_{2}^{+}}}\sum_{{k_{2} \in \mathcal{I}_{2}^{+}}} Y_{4}\times Y_{5}\}\approx\sigma_{W}^{2}\left(1-\cos(\MyTheta)\right)\sum_{k \in \mathcal{I}_{2}^{+}}|X_{q,k}|^{2}|H_{q,k}|^{2}\lambda_{2}^{2}(q,k)\label{equ:Y4Y5} 
\end{align}
\hrulefill
\vspace*{4pt}
\end{figure*}

Note that, \begin{math}
\Var\{\Im\left[\Delta\Theta_{q,\boldsymbol\theta}\right]\}=\E\{(\Im\left[\Delta\Theta_{q,\boldsymbol\theta}\right])^{2}\}-(\E\{\Im\left[\Delta\Theta_{q,\boldsymbol\theta}\right]\})^{2}
\end{math} where 
\begin{equation}
(\E\{\Im\left[\Delta\Theta_{q,\boldsymbol\theta}\right]\})^{2}=\sin^{2}(\MyTheta)\sigma_{W}^{4}\left(\sum_{{k\in \mathcal{I}_{2}^{+}}}\lambda_{2}(q,k)\right)^{2}
\end{equation}

Expanding $(\Im\left[\Delta\Theta_{q,\boldsymbol\theta}\right])^{2}$ into the form shown as \eqref{equ:expandingIm} with $\lambda_{3}(q,k)$ shown as \eqref{equ:lambda3}. $\E\{\sum_{{k_{1} \in \mathcal{I}_{2}^{+}}}\sum_{{k_{2}\in \mathcal{I}_{2}^{+}}} Y_{1}\times Y_{5}\}$ and $\E \{\sum_{{k_{1} \in \mathcal{I}_{2}^{+}}}\sum_{{k_{2} \in \mathcal{I}_{2}^{+}}} Y_{4}\times Y_{5}\}$ can be computed as \eqref{equ:Y1Y5} and \eqref{equ:Y4Y5}, while $\E \{\sum_{{k_{1} \in \mathcal{I}_{2}^{+}}}\sum_{{k_{2} \in \mathcal{I}_{2}^{+}}} Y_{2}\times Y_{5}\}$ and $\E \{\sum_{{k_{1} \in \mathcal{I}_{2}^{+}}}\sum_{{k_{2} \in \mathcal{I}_{2}^{+}}} Y_{3}\times Y_{5}\}$ can be neglected. $\Var\{\Im\left[\Delta\Theta_{q,\boldsymbol\theta}\right]\}$ under $\Athree$ and $\Afour$ is expressed by
\begin{equation}
\Var\{\Im\left[\Delta\Theta_{q,\boldsymbol\theta}\right]\}\approx \Xi N \frac{\SNR\left(1-\cos(\Theta_{q,\boldsymbol\theta})\right)}{(2+\SNR^{-1})^{2}}
\end{equation}
Substituting $\Var\{\Im\left[\Delta\Theta_{q,\boldsymbol\theta}\right]\}$ and $\left(\E\{\Re\left[\Delta\Theta_{q,\boldsymbol\theta}\right]\}\right)^{2}$ back into \eqref{equ:Varequ} yields 
\begin{equation}\label{equ:finalvar}
\Var\{\Delta\Theta_{q,\boldsymbol\theta}\} \approx \frac{1-\cos(\Theta_{q,\boldsymbol\theta})}{\Xi N \times \SNR}
\end{equation}
Therefore, 
\begin{equation}
\Var\{\widehat{\boldsymbol\theta_{\mathrm{DA}}}\}=\Var\{\underline{\mathbf{V}}\}=\Var\{(\mathbf{E}^{T}\mathbf{E})^{-1}\mathbf{E}^{T}\mathbf{V}\}
\end{equation}
Lengthy calculations lead to \eqref{equ:vareps} and \eqref{equ:vareta}. Under $\Athree$ and $\Afour$, the DA estimator reduces to NDA. 



\bibliographystyle{IEEEtranTCOM}
\bibliography{IEEEabrv,TCOM}
\end{document}